\documentstyle[12pt,epsf]{article}
\def\largelinestretch{\renewcommand{\baselinestretch}{1.3}}
\def\largelinestretch{\renewcommand{\baselinestretch}{1.3}}
\largelinestretch\small\normalsize

\title{
\bf Studies of Angular Correlations
    in the Decays $B^0_s\to J/\psi\,\phi$
    by Using the SIMUB Generator}

\vspace*{-5mm}
\author{A.~Bel'kov${}^1$, S.~Shulga${}^2$
\\[1ex]
\small
${}^1$
        Particle Physics Laboratory, Joint Institute for Nuclear
        Research,
\hfill\\[-2mm]
\small
        141980 Dubna, Moscow region, Russia
\hfill\\[-2mm]
\small
${}^2$
Francisk Skarina Gomel State University, Gomel, Belarus
\hfill\\[-3mm]
}
\date{}
\begin{document}

%\largelinestretch\normalsize
\thispagestyle{empty}
\begin{titlepage}
\thispagestyle{empty}
\maketitle
\begin{abstract}
\end{abstract}
%\normalsize
\vspace*{-10mm}
   The performance of the method of angular moments on the $\Delta\Gamma_s$ 
determination from analysis of untagged decays
$B^0_s(t),\overline{B}^0_s(t)\to J/\psi (\to l^+l^-)\,\phi (\to K^+K^-)$ is
examined by using the SIMUB generator.
   The results of Monte Carlo studies with evaluation of measurement 
errors are presented.
   The method of angular moments gives stable results for the estimate of
$\Delta\Gamma_s$ and is found to be an efficient and flexible tool for the
quantitative investigation of the $B^0_s\to J/\psi\,\phi$ decay.
   The statistical error of the ratio $\Delta\Gamma_s/\Gamma_s$ for values of 
this ratio in the interval [0.03, 0.3] was found to be independent on this 
value, being 0.015 for $10^5$ events. 

\vspace{0.5em}\noindent


\vspace{10mm}
\hfill
\end{titlepage}

\def\largelinestretch{\renewcommand{\baselinestretch}{1.6}}
\largelinestretch\small\normalsize

%-----------------------------------------------------------------------------
\section{Introduction}
%-----------------------------------------------------------------------------

   The study of decays 
$B^0_s(t),\overline{B}^0_s(t)\to J/\psi (\to l^+l^-)\,\phi (\to K^+K^-)$, 
which is one of the gold plated channels for $B$-physics studies at the LHC, 
looks very interesting from the physics point of view.
   It presents several advantages related to the dynamics of these decays,
characterized by proper-time-dependent angular distributions, which can 
be described in terms of bilinear combinations of transversity amplitudes.
   Their time evolution involves, besides the values of two transversity 
amplitudes at the proper time $t=0$ and their relative strong phases, the 
following fundamental parameters: the difference and average value of decay 
rates of heavy and light mass eigenstates of $B^0_s$ meson, $\Delta\Gamma_s$ 
and $\Gamma_s$, respectively, their mass difference $\Delta M_s$, and the 
CP-violating weak phase $\phi_c^{(s)}$.
   The angular analysis of the decays 
$B^0_s(t),\overline{B}^0_s(t)\to J/\psi (\to l^+l^-)\,\phi (\to K^+K^-)$ 
provides complete determination of the transversity amplitudes and, in 
principle, gives the access to all these parameters.

   In the present paper we examine the performance of the angular-moments 
method \cite{dighe2} applied to the angular analysis of untagged decays
$B^0_s(t),\overline{B}^0_s(t)\to J/\psi (\to l^+l^-)\,\phi (\to K^+K^-)$ for 
the determination of $\Delta\Gamma_s$.
   After giving the physics motivation in Section 2, we describe in the next 
section the method of angular moments based on weighting functions introduced 
in Ref.~\cite{dighe2}.
   For the case of $\Delta\Gamma_s$ determination this method is properly 
modified in Section 4. 
   The SIMUB-package \cite{SIMUB} for physics simulation of $B$-meson 
production and decays has been used for Monte Carlo studies. 
   A general information about the SIMUB package one can find in Appendix.
   The dynamics of the decays 
$B^0_s(t),\overline{B}^0_s(t)\to J/\psi (\to l^+l^-)\,\phi (\to K^+K^-)$  
is described by four-dimensional probability density functions depending on 
decay time and three physical angles.
   The algorithms of multidimensional random number generation have been
elaborated and then implemented in the package SIMUB to provide tools for 
Monte Carlo simulation of sequential two-body decays 
$B^0_s(t),\overline{B}^0_s(t)\to J/\psi (\to l^+l^-)\,\phi (\to K^+K^-)$  
in accordance with theoretical time-dependent angular distributions.
   This algorithms is discussed in Section 5.
   In Section 6 we present the results of the Monte Carlo studies of the 
angular analysis of the decays
$B^0_s(t),\overline{B}^0_s(t)\to J/\psi (\to l^+l^-)\,\phi (\to K^+K^-)$
with two sets of weighting functions for angular-moments method and 
concentrate on the evaluation of measurement errors and their dependence on 
statistics.

%-----------------------------------------------------------------------------
\section{Phenomenological description of the decays\\
     $B^0_s(t),\overline{B}^0_s(t)\to J/\psi (\to l^+l^-)\,\phi (\to K^+K^-)$}
%-----------------------------------------------------------------------------

   The angular distributions for decays 
$B^0_s(t),\overline{B}^0_s(t)\to J/\psi (\to l^+l^-)\,\phi (\to K^+K^-)$ are 
governed by spin-angular correlations (see \cite{jacob}-\cite{kutschke}) and 
involve three physically determined angles.
   In case of the so-called helicity frame~\cite{kramer1}, which is used in 
the present paper, these angles are defined as follows 
(see Fig.~1): 
\begin{itemize}
\item The $z$-axis is defined to be the direction of $\phi$-particle in the 
      rest frame of the $B^0_s$. The $x$-axis is defined as any arbitrary
      fixed direction in the plane normal to the $z$-axis. The $y$-axis is 
      then fixed uniquely via $y=z\times x$ (right-handed coordinate system).
\item The angles ($\Theta_{l^+}$, $\chi_{l^+}$) specify the direction of the 
      $l^+$ in the $J/\psi$ rest frame while ($\Theta_{K^+}$, $\chi_{K^+}$) 
      give the direction of $K^+$ in the $\phi$ rest frame. Since the 
      orientation of the $x$-axis is a matter of convention, only the 
      difference $\chi = \chi_{l^+}-\chi_{K^+}$ of the two azimuthal angles is
      physically meaningful.
\end{itemize}

    In the most general form the angular distribution for the decay 
$B^0_s(t)\to J/\psi (\to l^+l^-)\,\phi (\to K^+K^-)$ in case of a
{\it tagged $B^0_s$ sample} can be expressed as
\begin{equation}
\frac{d^4 N^{tag}(B^0_s)}
     {d\mbox{cos}\Theta_{l^+}\, d\mbox{cos}\Theta_{K^+}\, d\chi\, dt}
= \frac{9}{32\pi}\,\sum^6_{i=1} {\cal O}_i(t)
  g_i(\Theta_{l^+},\Theta_{K^+}, \chi)\,.
\label{angle_dist_tag1}
\end{equation}
   Here ${\cal O}_i$ ($i=1,...,6$) are time-dependent bilinear combinations of
the transversity amplitudes $A_0(t)$, $A_{||}(t)$ and $A_\bot(t)$ for the weak
transition $B^0_s(t)\to J/\psi\,\phi$ \cite{dighe1} (we treat these 
combinations as observables):
\begin{eqnarray}
&& {\cal O}_1=|A_0(t)|^2\,,\quad {\cal O}_2=|A_{||}(t)|^2\,,\quad 
   {\cal O}_3=|A_\bot(t)|^2\,,
\nonumber \\
&& {\cal O}_4=\mbox{Im}\Big( A_{||}^*(t) A_\bot(t)\Big)\,,\quad 
   {\cal O}_5=\mbox{Re}\Big( A_0^*(t) A_{||}(t)\Big)\,,\quad
   {\cal O}_6=\mbox{Im}\Big(A_0^*(t)A_\bot(t)\Big)\,,
\label{observ}
\end{eqnarray}
and the $g_i$ are functions of the angles $\Theta_{l^+}$, $\Theta_{K^+}$, 
$\chi$ only \cite{kramer1}:
\begin{eqnarray}
&& g_1=2\mbox{cos}^2\Theta_{K^+} \mbox{sin}^2\Theta_{l^+}\,,
\nonumber\\
&& g_2=\mbox{sin}^2\Theta_{K^+}(1-\mbox{sin}^2\Theta_{l^+}\mbox{cos}^2\chi)\,,
\nonumber\\
&& g_3=\mbox{sin}^2\Theta_{K^+}(1-\mbox{sin}^2\Theta_{l^+}\mbox{sin}^2\chi)\,,
\nonumber\\
&& g_4=-\mbox{sin}^2\Theta_{K^+}\mbox{sin}^2\Theta_{l^+}\mbox{sin}2\chi\,,
\nonumber\\
&& g_5=\frac{1}{\sqrt{2}}\mbox{sin}2\Theta_{l^+}\mbox{sin}2\Theta_{K^+}
                         \mbox{cos}\chi\,,
\nonumber\\
&& g_6=\frac{1}{\sqrt{2}}\mbox{sin}2\Theta_{l^+}\mbox{sin}2\Theta_{K^+}
                         \mbox{sin}\chi\,.
\label{g_VllVpp}
\end{eqnarray}
   For the decay 
$\overline{B}^0_s(t)\to J/\psi (\to l^+l^-)\,\phi (\to K^+K^-)$ in case of a 
{\it tagged $\overline{B}^0_s$ sample} the angular distribution is given by 
\begin{equation}
\frac{d^4 N^{tag}(\overline{B}^0_s)}
     {d\mbox{cos}\Theta_{l^+}d\mbox{cos}\Theta_{K^+}d\chi dt}
= \frac{9}{32\pi}\,\sum^6_{i=1} \overline{{\cal O}}_i(t)
  g_i(\Theta_{l^+},\Theta_{K^+}, \chi)
\label{angle_dist_tag2}
\end{equation}
with the same angular functions $g_i$ and
\begin{eqnarray}
&& \overline{{\cal O}}_1=|\bar{A}(t)|^2\,,\quad 
   \overline{{\cal O}}_2=|\bar{A}_{||}(t)|^2\,,\quad 
   \overline{{\cal O}}_3=|\bar{A}_\bot(t)|^2\,, 
\nonumber \\
&& \overline{{\cal O}}_4=\mbox{Im}\Big(\bar{A}_{||}^*(t)\bar{A}_\bot(t)\Big)\,,
   \quad
   \overline{{\cal O}}_5=\mbox{Re}\Big( \bar{A}_0^*(t) \bar{A}_{||}(t))\,,\quad
   \overline{{\cal O}}_6=\mbox{Im}\Big( \bar{A}_0^*(t)\bar{A}_\bot(t) \Big)\,,
\label{observ_bar}
\end{eqnarray}
where $\bar{A}_0(t)$, $\bar{A}_{||}(t)$ and $\bar{A}_\bot(t)$ are the 
transversity amplitudes for the transition 
$\overline{B}^0_s(t)\to J/\psi\,\phi$.

   The time dependence of the transversity amplitudes for the transitions
$B^0_s(t)\,,\overline{B}^0_s(t)\to J/\psi\, \phi$ is not of purely exponential 
form due to the presence of $B^0_s-\overline{B}^0_s$ mixing.
   This mixing arises due to either a mass difference or a decay-width 
difference between the mass eigenstates of the $(B^0_s-\overline{B}^0_s)$ 
system.
   The time evolution of the state $|B^0_s(t)\rangle$ of an initially, i.e. 
at time $t=0$, present $B^0_s$ meson can be described in general form as 
follows:
$$
|B^0_s(t)\rangle = g_+(t)|B^0_s\rangle 
                  +g_-(t)|\overline{B}^0_s\rangle\,,\qquad
g_+(t=0) = 1\,, \quad  g_-(t=0) = 0\,,
$$
i.e., the state $|B^0_s(t)\rangle$ at time $t$ is a mixture of the flavor
states $|B^0_s\rangle$ and $|\overline{B}^0_s\rangle$ with probabilities 
defined by the functions $g_+(t)$ and $g_-(t)$.
   In analogous way, the time evolution of the state 
$|\overline{B}^0_s(t)\rangle$ of an initially present $\overline{B}^0_s$ meson
is described by the relation
$$
|\overline{B}^0_s(t)\rangle = \bar{g}_+(t)|B^0_s\rangle 
                             +\bar{g}_-(t)|\overline{B}^0_s\rangle\,,\qquad
\bar{g}_+(t=0) = 0\,, \quad \bar{g}_-(t=0) = 1\,.
$$

   Diagonalization of the full Hamiltonian (see \cite{richman1} for more 
details) gives 
\begin{eqnarray}
&& g_+(t) = \frac{1}{2}\Big( e^{-i\mu_L t} +  e^{-i\mu_H t}\Big)\,,\quad 
 g_-(t) = \frac{\alpha}{2}\Big( e^{-i\mu_L t} -  e^{-i\mu_H t}\Big)\,,
\nonumber\\
&& \bar{g}_+(t) = g_-(t)/\alpha^2\,,\qquad\qquad\quad~ 
   \bar{g}_-(t) = g_+(t)\,. 
\label{g_functions}
\end{eqnarray} 
   Here $\mu_{L/H} \equiv M_{L/H}-(i/2)\Gamma_{L/H}$ are eigenvalues of the 
full Hamiltonian corresponding to the masses and total widths of ``light'' and
 ``heavy'' eigenstates $|B_{L/H}\rangle$, and $\alpha$ is a phase factor 
defining the $CP$ transformation of flavor eigenstates of the neutral 
$B_s$-meson system: 
$CP|B^0_s\rangle  = \alpha |\overline{B}^0_s\rangle$.
   In the case $|\alpha| \not= 1$ the probability for $B^0_s$ to oscillate to 
a $\overline{B}^0_s$ is not equal to the probability of a $\overline{B}^0_s$ 
to oscillate to a $B^0_s$.
   Such an asymmetry in mixing is often referred to as {\it indirect} $CP$ 
violation, which is negligibly small in case of the neutral $B$-meson system.

   The time evolution of the transversity amplitudes $A_f(t)~ (f=0,||,\bot)$
is given by the equations 
\begin{equation}
\nonumber
A_f(t) = A_f(0) 
           \bigg[g_+(t) + g_-(t)\frac{1}{\eta_{CP}^f\alpha}\xi_f^{(s)}\bigg]\,,
\quad
\bar{A}_f(t) = A_f(0) 
                 \bigg[ \bar{g}_+(t) 
                       +\bar{g}_-(t)\frac{1}{\eta_{CP}^f\alpha}\xi_f^{(s)}
                 \bigg]\,.
\label{time_evol_transit}
\end{equation}
   Here $\eta_{CP}^f$ are eigenvalues of CP-operator acting on the 
transversity components of the final state which are eigenstates of 
$CP$-operator:
\begin{eqnarray*}
&&  CP|J/\psi\,\phi\rangle_f = \eta_{CP}^f|J/\psi\,\phi\rangle_f\,,\quad 
    (f = 0,||,\bot)\,,
\nonumber \\
&&  \eta_{CP}^0 =1\,,\quad \eta_{CP}^{||} =1\,,\quad \eta_{CP}^\bot = -1\,,  
\end{eqnarray*}
and $\xi_f^{(s)}$ is the CP-violating weak phase \cite{fleischer1}:
%\begin{equation}
$$
\xi_f^{(s)}  = e^{-i\phi_c^{(s)}}\,,\quad
\phi_c^{(s)} =  2[\mbox{arg}(V_{ts}^*V_{tb})-\mbox{arg}(V_{cq}^*V_{cb})]
             = -2\delta\gamma\,,
$$
%\label{weak-phase}
%\end{equation}
where $\delta$ is the complex phase in the standard parameterization of the 
CKM matrix elements $V_{ij}$ ($i\in \{u,c,t\}$, $j\in \{d,s,b\}$), and 
$\gamma$ is the third angle of the unitarity triangle.

   The phase $\phi_c^{(s)}$ is very small and vanishes at leading order in
the Wolfenstein expansion.
   Taking into account higher-order terms in the Wolfenstein parameter
$\lambda = \mbox{sin} \theta_C = 0.22$ gives a non-vanishing result
\cite{dunietz3}:
$$
\phi_c^{(s)} =  -2\lambda^2\eta = -2\lambda^2 R_b \sin \gamma\,.
$$
   Here 
$$
R_b \equiv \frac{1}{\lambda}\frac{|V_{ub}|}{|V_{cb}|}
$$
is constrained by present experimental data as $R_b = 0.36\pm 0.08$
\cite{Rb_exp}.
   Using the estimate $\gamma = (59\pm 13)^o$ \cite{PDG}, the
following constrain can be obtained for the phase $\phi_c^{(s)}$:
\begin{equation}
\phi_c^{(s)} = -0.03\pm 0.01\,.
\label{weak-phase_ccs}
\end{equation}

   According to Eq.~(\ref{time_evol_transit}) at time $t=0$, the transversity
amplitudes of $B^0_s\,,\overline{B^0}_s\to J/\psi\,\phi$ decays depend on the 
same observables $|A_0(0)|$, $|A_{||}(0)|$, $|A_\bot(0)|$ and on the two 
CP-conserving strong phases, $\delta_1 \equiv arg[A_{||}^*(0) A_\bot(0)]$ and 
$\delta_2 \equiv arg[A_0^*(0)A_\bot(0)]$.
   Time-reversal invariance of strong interactions forces the form factors 
parameterizing quark currents to be all relatively real and, consequently, 
naive factorization leads to the following common properties of the 
observables:
$$
\mbox{Im}[A_0^*(0)A_\bot(0)]     = 0\,,\quad
\mbox{Im}[ A_{||}^*(0)A_\bot(0)] = 0\,,\quad
\mbox{Re}[A_0^*(0)A_{||}(0)]     = \pm |A_0(0) A_{||}(0)|\,.
$$
   Moreover, in the absence of strong final-state interactions, 
$\delta_1 = \pi$ and $\delta_2 =0$.

   In the framework of the effective Hamiltonian approach the two body decays,
both $B^0_s\to J/\psi\,\phi$ and $B^0_d\to J/\psi\,K^\star$, correspond to the 
transitions $\bar{b}\to \bar{s}\bar{c}c$ with topologies of color-suppressed 
spectator diagrams shown in Fig.~2.
   Factorizing the hadronic matrix elements of the four-quark operators of the
effective Hamiltonian into hadronic matrix elements of quark currents, the
transversity amplitudes $|A_0(0)|$, $|A_{||}(0)|$, $|A_\bot(0)|$ of decays  
${B}^0_q, \overline{B}^0_q\to J/\psi V ((q,V)\in \{ (s,\phi), (d,K^\star )\})$
can be expressed in terms of effective Wilson coefficient functions, constants
of $J/\psi$ decay, and form factors of transitions $B_q\to V$ induced by quark
currents \cite{dighe2}.
   In Table~\ref{tab:observ_theor} we collect the predictions of 
Ref.~\cite{dighe2} for the transversity amplitudes of $B^0_s\to J/\psi\,\phi$ 
($B^0_d\to J/\psi K^\star$) calculated with $B\to K^\star$ form factors 
given by different models \cite{wirbel,soares,cheng}.
   The $B\to K^\star$ form factors can be related to the $B\to \phi$ case by
using SU(3) flavor symmetry.
   The most precise polarization measurements performed recently in
decays $B\to J/\psi\,K^\star$:
\begin{eqnarray*}
&& |A_0(0)|^2=0.60\pm 0.04\,,\quad |A_\bot (0)|^2=0.16\pm 0.03\quad
   \mbox{(BaBar~~\cite{BaBar})}\,,\\
&& |A_0(0)|^2=0.62\pm 0.04\,,\quad |A_\bot (0)|^2=0.19\pm 0.04\quad
   \mbox{(Belle~~\cite{Belle})}\,,
\end{eqnarray*}
confirm the predictions based on the model \cite{cheng}.

%--------------------------------
\section{Angular-moments method}
%--------------------------------

   The angular distributions for decays
$B^0_s(t),\overline{B}^0_s(t)\to J/\psi (\to l^+l^-)\,\phi (\to K^+K^-)$ in case
of tagged $B^0_s$ and $\overline{B}^0_s(t)$ samples 
(see Eqs.~(\ref{angle_dist_tag1}) and (\ref{angle_dist_tag2}), respectively)
as well as in case of the untagged sample can be expressed in the most general 
form in terms of observables $b_i(t)$:
\begin{equation}
f_0(\Theta_{l^+},\Theta_{K^+},\chi;t) = \frac{9}{32\pi}\,\sum^6_{i=1} b_i(t)
                                      g_i(\Theta_{l^+},\Theta_{K^+},\chi)\,.
\label{angle_dist}
\end{equation}
   The explicit time dependence of observables is given by the following
relations:
\begin{eqnarray}
&& b_1(t) = |A_0(0)|^2 G_L(t)\,,  
\nonumber \\
&& b_2(t) = |A_{||}(0)|^2 G_L(t)\,,
\nonumber \\
&& b_3(t) = |A_\bot (0)|^2 G_H(t)\,,
\nonumber \\
&& b_4(t) = |A_{||}(0)|\,|A_\bot (0)|\,Z_1(t)\,,
\nonumber \\
&& b_5(t) = |A_{0}(0)|\,|A_{||}(0)\,|G_L(t)\,\mbox{cos}(\delta_2-\delta_1)\,,
\nonumber \\
&& b_6(t) = |A_0(0)|\,|A_\bot (0)|\,Z_2(t)\,,
\label{observ-time-depend}
\end{eqnarray}
where we have used the general compact 
notations: 
\begin{eqnarray*}
&& G_{L/H}(t) = \frac{1}{2}
\Big[(1\pm\mbox{cos}\phi_c^{(s)})\mbox{e}^{-\Gamma_L t }+
     (1\mp\mbox{cos}\phi_c^{(s)})\mbox{e}^{-\Gamma_H t }
\Big] \,,
\nonumber \\
&& Z_{1,2}(t) = \frac{1}{2}
\Big(\mbox{e}^{-\Gamma_H t}- \mbox{e}^{-\Gamma_L t}
\Big)\mbox{cos}\delta_{1,2}\mbox{sin}\phi_c^{(s)}
\end{eqnarray*}
-- for observables $b_i\equiv ({\cal O}_i+\overline{{\cal O}}_i)/2$ 
in case of the untagged sample with equal initial numbers of $B^0_s$ and 
$\overline{B}^0_s$, while
\begin{eqnarray*}
&& G^{(B^0_s)/(\overline{B}^0_s)}_{L/H}(t) = G_{L/H}(t) 
                     \pm\mbox{e}^{-\Gamma_s t}\,\mbox{sin}(\Delta M t)
                     \,\mbox{sin}\phi_c^{(s)} \,,
\nonumber \\
&& Z_{1,2}^{(B^0_s)/(\overline{B}^0_s)}(t) = Z_{1,2}(t) 
\pm\,\mbox{e}^{-\Gamma_s t}\,\Big[
           \mbox{sin}\delta_{1,2}\mbox{cos}(\Delta M t)
          -\mbox{cos}\delta_{1,2}\mbox{sin}(\Delta M t)\mbox{sin}\phi_c^{(s)}
                           \Big]
\end{eqnarray*}
-- for observables $b_i^{(B^0_s)}\equiv {\cal O}_i$ and 
$b_i^{(\overline{B}^0_s)}\equiv \overline{{\cal O}}_i$ in case of tagged 
$B^0_s$ and $\overline{B}^0_s(t)$ samples, respectively, with 
$\Gamma_s \equiv (\Gamma_L + \Gamma_H)/2$.
   It is easy to see that both in the tagged and untagged case we have
$$
  G_{L/H}(t)|_{\phi_c^{(s)}=0} = \mbox{e}^{-\Gamma_{L/H}t}\,.
$$

   According to Ref.~\cite{dighe2}, the observables $b_i(t)$ can be extracted
from distribution function~(\ref{angle_dist}) by means of weighting functions
$w_i(\Theta_{l^+}, \, \Theta_{K^+}, \, \chi )$ for each $i$ such that
\begin{equation}
\frac{9}{32\pi}\int  
                 d\mbox{cos}\Theta_{l^+}
                 d\mbox{cos}\Theta_{K^+}
                 d\chi\,\, w_i(\Theta_{l^+},\Theta_{K^+},\chi ) \,\, 
                 g_j(\Theta_{l^+},\Theta_{K^+},\chi ) = \delta_{ij} \,,
\label{ort_cond}
\end{equation}
projecting out the desired observable alone:
\begin{equation}
b_i(t) = \int d\mbox{cos}\Theta_{l^+} d\mbox{cos}\Theta_{K^+} d\chi\,\, 
              f_0(\Theta_{l^+},\Theta_{K^+},\chi\,;t )\,\, 
              w_i(\Theta_{l^+},\Theta_{K^+},\chi )\,.
\label{project_observ}
\end{equation}
   The angular-distribution function (\ref{angle_dist}) obeys the condition
\begin{equation}
L(t) \equiv \int d\mbox{cos}\Theta_{l^+} d\mbox{cos}\Theta_{K^+} d\chi\,\, 
                 f_0(\Theta_{l^+},\Theta_{K^+},\chi\,;t )\,\,
     = b_1(t)+b_2(t)+b_3(t)\,.
\label{L_norm}
\end{equation}

   For decays $B\to J/\psi (\to l^+l^-)\,\phi (\to K^+K^-)$, the explicit 
expressions of weighting functions, given in Table~5 of Ref.~\cite{dighe2} for
physically meaningful angles in the transversity frame, get the following 
form (Set A) after transformation into the helicity frame:
\begin{eqnarray}
&& w^{(A)}_1 = 2 - 5 \cos^2{\Theta_{l^+}}\,,
\nonumber \\
&& w^{(A)}_2 = 2 - 5 \sin^2{\Theta_{l^+}}\cos^2{\chi}\,,
\nonumber \\
&& w^{(A)}_3 = 2 - 5 \sin^2{\Theta_{l^+}}\sin^2{\chi}\,,
\nonumber \\
&& w^{(A)}_4 =-\frac{5}{2}\sin^2{\Theta_{K^+}}\sin{2\chi}\,,
\nonumber \\
&& w^{(A)}_5 = \frac{25}{4\sqrt{2}} 
               \sin{2\Theta_{K^+}}\sin{2\Theta_{l^+}}\cos{\chi}\,,
\nonumber \\
&& w^{(A)}_6 = \frac{25}{4\sqrt{2}}
               \sin{2\Theta_{K^+}}\sin{2\Theta_{l^+}}\sin{\chi}\,.
\label{set_A}
\end{eqnarray}

   The expressions of Eq.~(\ref{set_A}) are not unique and there are many 
legitimate choices of weighting functions.
   A particular set can be derived by linear combination of angular functions 
$g_i$ (see \cite{dighe2} for more discussions):
\begin{equation}
\label{eq:w_sum_g}
w_i(\Theta_{l^+},\Theta_{K^+},\chi) = \sum_{j=1}^6 \lambda_{ij}
g_j(\Theta_{l^+},\Theta_{K^+},\chi)\,,
\end{equation}
where the 36 unknown coefficients $\lambda_{ij}$ are solutions of 36 equations
\begin{eqnarray}
\frac{9}{32\pi} \sum_{j=1}^6 \lambda_{ij} 
\int d\mbox{cos}\Theta_{l^+}\, d\mbox{cos}\Theta_{K^+}\, d\chi \,\,
     g_j(\Theta_{l^+},\Theta_{K^+},\chi)\,
     g_k (\Theta_{l^+},\Theta_{K^+},\chi) = 
                                            \delta_{ik}\,.
\end{eqnarray}
   The weighting functions (set B) corresponding to the linear combination of 
the angular functions (\ref{g_VllVpp}) are given by
\begin{eqnarray}
w^{(B)}_1 &=& \frac{1}{12}[28\cos^2{\Theta_{K^+}}\sin^2{\Theta_{l^+}} -
                           3\sin^2{\Theta_{K^+}}(1+\cos^2{\Theta_{l^+}})]\,,
\nonumber \\
w^{(B)}_2 &=& -\,\frac{1}{8}[4\cos^2{\Theta_{K^+}}\sin^2{\Theta_{l^+}}
              -29 \sin^2{\Theta_{K^+}}(1-\sin^2{\Theta_{l^+}}\cos^2{\chi}) 
\nonumber \\ &&~~~~
              +21\sin^2{\Theta_{K^+}}(1-\sin^2{\Theta_{l^+}}\sin^2{\chi})]\,,
\nonumber \\
w^{(B)}_3 &=& -\,\frac{1}{8}[4\cos^2{\Theta_{K^+}}\sin^2{\Theta_{l^+}} 
              +21\sin^2{\Theta_{K^+}}(1-\sin^2{\Theta_{l^+}}\cos^2{\chi}) 
\nonumber \\ &&~~~~
              -29\sin^2{\Theta_{K^+}}(1-\sin^2{\Theta_{l^+}}\sin^2{\chi})]\,,
\nonumber \\
w^{(B)}_4 &=& -\,\frac{25}{8}\sin^2{\Theta_{K^+}}
                           \sin^2{\Theta_{l^+}}\sin{2\chi}\,,
\nonumber \\
w^{(B)}_5  &=& w^{(A)}_5\,,
\nonumber \\
w^{(B)}_6 &=& w^{(A)}_6\,.
\label{set_B}
\end{eqnarray}

   For a limited number of experimental events $N$ in the time bin around the 
fixed value of the proper time $t$, distributed according to the angular 
function (\ref{angle_dist}), it is convenient to introduce the normalized 
observables
\begin{equation}
 \bar{b}_i(t) \equiv b_i(t)/L(t)
\label{project_observ-barb}
\end{equation}
with normalization factor $L(t)$ given by Eq.~(\ref{L_norm}).
   Then, as it follows from the  Eq.~(\ref{project_observ}), the observables
$\bar{b}_i(t)$~(\ref{project_observ-barb}) are measured experimentally by
\begin{equation}
\bar{b}_i^{(exp)} = \frac{1}{N}\sum_{j=1}^{N}\,w^j_i
\label{observ_exp}
\end{equation}
with summation over events in a time bin around $t$.
   Here $w^j_i \equiv w_i(\Theta^j_{l^+},\Theta^j_{K^+},\chi^j)$, where 
$\Theta^j_{l^+}$, $\Theta^j_{K^+}$ and $\chi^j$ are angles measured in the 
$j$-th event.
   The statistical measurement error of the observable 
(\ref{observ_exp}) can be estimated as
$$
\delta \bar{b}_i^{(exp)} = \frac{1}{N}\sqrt{\sum_{j=1}^{N}
                        (\bar{b}_i^{(exp)}- w_i^j)^2}\,,
$$
with summation over all events in the same time bin.

%--------------------------------------
 \section{Time-integrated observables}
%--------------------------------------

   For data analysis it is rather convenient to use the time-integrated 
observables \cite{DGamma} defined as
\begin{equation}
\tilde{b}_i(T_0) = \frac{1}{\tilde{L}(T)}\,
\int_0^{T_0} dt\int d\mbox{cos}\Theta_{l^+}\, d\mbox{cos}\Theta_{K^+}\, d\chi 
                     \,w_i(\Theta_{l^+},\Theta_{K^+},\chi)
                 \,\,f_0(\Theta_{l^+},\Theta_{K^+},\chi;t)
\label{observ_tild}
\end{equation}
with argument $T_0 \le T$, where $T$ is the maximal value of the $B$-meson 
proper time measured for the sample of events being used, and $\tilde{L}(T)$ 
is a new normalization factor, which has the form:
\begin{eqnarray}
\tilde{L}(T) 
&\equiv & \int_0^{T} L(t) =
\int_0^{T} dt\int d\mbox{cos}\Theta_{l^+}\, d\mbox{cos}\Theta_{K^+}\, d\chi
                  \,\,f_0(\mbox{cos}\Theta_{l^+},
                              \mbox{cos}\Theta_{K^+},\chi;t)\,\, =
\nonumber \\ 
&=& (|A_0(0)|^2 + |A_{||}(0)|^2)\,\tilde{G}_L(T) + |A_\bot (0)|^2\,\tilde{G}_H(T)\,,
\label{norm-fact}
\end{eqnarray}
where, in the compact notations used in Eq.~(\ref{observ-time-depend}),
$$
\tilde{G}_{L/H}(T) \equiv \int_0^T dt\, G_{L/H}(t)\,.
$$
   The following normalization condition is valid for the observables
(\ref{observ_tild}): $\tilde{b}_1+\tilde{b}_2+\tilde{b}_3=1$.
   For a limited number of experimental events $N(T)$, measured in the 
proper time region $t\in [0,T]$, Eq.~(\ref{observ_tild}) reduces to
\begin{equation}
\tilde{b}_i^{(exp)}(T_0) = \frac{1}{N(T)}\sum_{j=1}^{N(T_0)}\,w_i^j
\label{observ_tilde_exp}
\end{equation}
with summation over all events $N(T_0)$ in the time interval $t\in [0,T_0]$
for $T_0 \le T$.
   In case of the untagged sample we have
\begin{equation}
\tilde{G}_{L/H}(T) = -\frac{1}{2}\bigg[
      (1\pm\mbox{cos}\phi_c^{(s)})\frac{\mbox{e}^{-\Gamma_L T}-1}{\Gamma_L} 
    + (1\mp\mbox{cos}\phi_c^{(s)})\frac{\mbox{e}^{-\Gamma_H T}-1}{\Gamma_H}
                                           \bigg]
\label{G_tilde}
\end{equation}
and
\begin{eqnarray*}
\tilde{Z}(T) &\equiv & 
                    \frac{1}{\mbox{cos}\delta_{1,2}\,\mbox{sin}\phi_c^{(s)}}
                    \int_0^T dt\,{Z}_{1,2}(T)
\nonumber \\
& = & -\frac{1}{2}\,\bigg[(\mbox{e}^{-\Gamma_H T}-1)/\Gamma_H 
                        -(\mbox{e}^{-\Gamma_L T}-1)/\Gamma_L\bigg]\,.
\end{eqnarray*}

   For the untagged sample the explicit form of time-integrated normalized 
observables (\ref{observ_tild}) in terms of the functions 
$\tilde{G}_{L/H}(T)$ and $\tilde{Z}(T)$ is given by
\begin{eqnarray}
&&
\tilde{b}_1(T_0)=|A_0(0)|^2\,\tilde{G}_L(T_0)/\tilde{L}(T)\,,
\nonumber \\ && 
\tilde{b}_2(T_0)=|A_{||}(0)|^2\,\tilde{G}_L(T_0)/\tilde{L}(T)\,,
\nonumber \\ && 
\tilde{b}_3(T_0)=|A_{\bot}(0)|^2\,\tilde{G}_H(T_0)/\tilde{L}(T)\,,
\nonumber \\ && 
\tilde{b}_4(T_0)=|A_{||}(0)|\,|A_{\bot}(0)|\,\tilde{Z}(T_0)\,
                          \mbox{cos}\delta_1\,\mbox{sin}\phi_c^{(s)}/\tilde{L}(T)\,,
\nonumber \\ && 
\tilde{b}_5(T_0)=|A_0(0)|\,|A_{||}(0)|\,\tilde{G}_L(T_0)\,
                          \mbox{cos}(\delta_2-\delta_1)/\tilde{L}(T)\,, 
\nonumber \\ && 
\tilde{b}_6(T_0)=|A_0(0)|\,|A_{\bot}(0)|\,\tilde{Z}(T_0)\,
                          \mbox{cos}\delta_2\,\mbox{sin}\phi_c^{(s)}/\tilde{L}(T)\,.
\label{observ_tild_ut_1}
\end{eqnarray}
   In the Standard Model (SM) $\mbox{sin}\phi_c^{(s)}\approx 0$ and the
observables $\tilde{b}_{4,6}(T_0)$ are vanishing.
   In case of a new physics signal the values of 
$\mbox{sin}\phi_c^{(s)}$ and $\tilde{b}_{4,6}(T_0)$ can be sizable, however.

   The following relations are valid for the observables 
(\ref{observ_tild_ut_1}):
\begin{eqnarray*}
&& \tilde{b}_4(T_0) = 
   \mbox{cos}\delta_1\,\mbox{sin}\phi_c^{(s)}\,\tilde{Z}(T_0)
   \sqrt{\frac{\tilde{b}_2(T_0)\,\tilde{b}_3(T_0)}
              {\tilde{G}_L(T_0)\,\tilde{G}_H(T_0)}}\,,
\nonumber \\
&& \tilde{b}_5(T_0) = 
   \mbox{cos}(\delta_2-\delta_1)
   \sqrt{\tilde{b}_1(T_0)\,\tilde{b}_2(T_0)}\,,
\nonumber \\
&& \tilde{b}_6(T_0) =   
   \mbox{cos}\delta_2\,\mbox{sin}\phi_c^{(s)}\,\tilde{Z}(T_0)
   \sqrt{\frac{\tilde{b}_1(T_0)\,\tilde{b}_3(T_0)}
              {\tilde{G}_L(T_0)\,\tilde{G}_H(T_0)}}\,.
\end{eqnarray*}

   If we introduce the function
\begin{eqnarray}
  \tilde{\gamma}(T) \equiv \tilde{G}_H(T)/\tilde{G}_L(T)\,,
\label{gamma}
\end{eqnarray}
then, the values of initial transversity amplitudes at $t=0$ and the
strong-phase difference $(\delta_2-\delta_1)$ are determined from the
observables $\tilde{b}_i(T) \equiv \tilde{b}_i(T=T_0)$ by 
\begin{eqnarray}
&& |A_0(0)|^2 = \frac{\tilde{b}_1(T)}
                {\tilde{b}_1(T)+\tilde{b}_2(T)
                +\tilde{b}_3(T)/\tilde{\gamma}(T)}\,,
\nonumber \\
&& |A_{||}(0)|^2 = \frac{\tilde{b}_2(T)}
                   {\tilde{b}_1(T)+\tilde{b}_2(T)
                   +\tilde{b}_3(T)/\tilde{\gamma}(T)}\,,
\nonumber \\
&& |A_{\bot}(0)|^2 = \frac{\tilde{b}_3(T)/\tilde{\gamma}(T)}
                     {\tilde{b}_1(T)+\tilde{b}_2(T)
                     +\tilde{b}_3(T)/\tilde{\gamma}(T)}\,,
\nonumber \\
&&  \mbox{cos}(\delta_2 - \delta_1) = 
    \frac{\tilde{b}_5(T)}
    {\sqrt{\tilde{b}_1(T)\,\tilde{b}_2(T)}}\,,
\label{observ_tild_ut_3}
\end{eqnarray}
where we consider the initial amplitudes normalized as 
$|A_0(0)|^2 + |A_{||}(0)|^2 + |A_\bot (0)|^2 = 1$.
   We have also:
\begin{equation}
\mbox{sin}\phi_c^{(s)}\,\mbox{cos}\delta_{1,2} =
          \frac{\tilde{b}_{4,6}(T)}
               {\sqrt{\tilde{b}_{2,1}(T)\tilde{b}_3(T)}}\,
          \frac{\sqrt{\tilde{G}_L(T)\,\tilde{G}_H(T)}}
               {\tilde{Z}(T)}\,.
\label{sin_phase}
\end{equation}

   For extraction of the $B^0_s$-width difference 
$\Delta \Gamma_s \equiv \Gamma_H - \Gamma_L$ from experimental data it is 
convenient to use a special set of the time-integrated normalized observables
\cite{DGamma}:
\begin{equation}
\hat{b}_i(T_0) = \frac{1}{\tilde{L}(T)}\,
\int_0^{T_0} dt\int d\mbox{cos}\Theta_{l^+} d\mbox{cos}\Theta_{K^+} d\chi\,
                  w_i(\Theta_{l^+},\Theta_{K^+},\chi)\,\,
                  \mbox{e}^{\Gamma^\prime t}\,
                  f_0(\Theta_{l^+},\Theta_{K^+},\chi;t)\,,
\label{observ_hat}
\end{equation}
where $\Gamma^\prime$ is some arbitrary initial approximation of the 
$B^0_s$-meson total decay width.
   These observables can be extracted from the experimental events $N(T)$, 
measured in the proper time region $t\in [0,T]$, by using the formula
\begin{equation}
\hat{b}_i^{(exp)}(T_0) = \frac{1}{N(T)}\sum_{j=1}^{N(T_0)}\,W^j_i\,,
\label{observ_hat_exp}
\end{equation}
where $W^j_i\equiv\mbox{e}^{\Gamma^\prime t^j}\,w_i^j$, and summation is 
performed over all events $N(T_0)$ in the time interval $t^j\in [0,T_0]$.

   For the untagged sample, the explicit expressions for the time-integrated 
observables (\ref{observ_hat}) can be easily obtained by replacing 
$\tilde{b}_i$, $\tilde{G}_{L/H}$ and $\tilde{Z}$ in the expressions of 
Eq.~(\ref{observ_tild_ut_1}) by $\hat{b}_i$, $\hat{G}_{L/H}$ and $\hat{Z}$, 
respectively, (with the same normalization factor (\ref{norm-fact})) after 
introducing the following notations
\begin{eqnarray}
\hat{G}_{L/H}(T) &\equiv& 
   \int_0^T dt\, \mbox{e}^{\Gamma^\prime t}\,G_{L/H}(t)
\nonumber \\
   &=& (1\pm\mbox{cos}\phi_c^{(s)})\,
     \frac{\mbox{e}^{ \Delta\Gamma_LT/2}-1}{\Delta\Gamma_L}
    -(1\mp\mbox{cos}\phi_c^{(s)})\,
     \frac{\mbox{e}^{-\Delta\Gamma_HT/2}-1}{\Delta\Gamma_H}\,,
\label{G_hat} \\
 \hat{Z}(T) &\equiv&  
                    \frac{1}{\mbox{cos}\delta_{1,2}\,\mbox{sin}\phi_c^{(s)}}
                    \int_0^T dt\,\mbox{e}^{\Gamma^\prime t}\,Z_{1,2}(T)
\nonumber \\
   &=& \frac{1-\mbox{e}^{ \Delta\Gamma_LT/2}}{\Delta\Gamma_L}
      +\frac{1-\mbox{e}^{-\Delta\Gamma_HT/2}}{\Delta\Gamma_H}\,.
\nonumber
\end{eqnarray}
where $\Delta\Gamma_{L/H}$ are auxiliary parameters given by
\begin{equation}
\Delta\Gamma_L =2(\Gamma^\prime -\Gamma_L)\,,\qquad
\Delta\Gamma_H =-2(\Gamma^\prime -\Gamma_H)\,.
\label{Delta_G_L/H}
\end{equation}
   Eq.~(\ref{observ_tild_ut_3}) is also valid after such a replacement.

%-----------------------------------------------------------------
 \section{Algorithms of multidimensional random number generation}
%-----------------------------------------------------------------

   The time-dependent angular distribution function (\ref{angle_dist}) is used
in the SIMUB generator \cite{SIMUB} (see also Appendix) as density of the
probability function for Monte Carlo simulation of the vertex and kinematics 
of the final-state particles in case of decays
$B^0_s(t),\overline{B}^0_s(t)\to J/\psi (\to l^+l^-)\,\phi (\to K^+K^-)$.
   The variables of the function $f_0(\Theta_{l^+},\Theta_{K^+},\chi;t)$
can not be factorized and randomly generated in an independent way.
   Nevertheless, the random generation of $\mbox{cos}\,\Theta_{l^+}$, 
$\mbox{cos}\,\Theta_{K^+}$, $\chi$ and $t$ can be performed either 
simultaneously according to the distribution function (\ref{angle_dist}) by 
using four-dimensional random generator or successively, one after 
another, by using single-dimensional random number generators with 
accordance to distribution functions obtained by successive integration of
the function (\ref{angle_dist}) over its variables.
   
   Let as consider the latter approach in the case of sequential random
generation of the variables $t$, $\chi$, $\mbox{cos}\,\Theta_{l^+}$ and
$\mbox{cos}\,\Theta_{K^+}$. 
   The following three distribution functions are used in this case:
\begin{eqnarray}
&&
f_1(\Theta_{l^+},\chi;t) \equiv \int_{-1}^{+1} d\mbox{cos}\Theta_{K^+}\,
                                  f_0(\Theta_{l^+},\Theta_{K^+},\chi;t)\,,
\nonumber\\ &&
f_2(\chi;t) \equiv \int_{-1}^{+1} d\mbox{cos}\Theta_{l^+}\,
                                  f_1(\Theta_{l^+},\chi;t)\,,
\label{angle_dist_1}
\end{eqnarray}
and
\begin{equation}
f_3(t) \equiv \int_{0}^{2\pi} d\chi f_2(\chi;t)\,
              = b_1(t)+b_2(t)+b_3(t)\,. 
\label{time_dist}
\end{equation}
   The functions of Eqs.~(\ref{angle_dist_1}) and (\ref{time_dist}) 
present the most important experimentally observable distributions.
   In case of sequential decays 
$B^0_s(t),\overline{B}^0_s(t)\to J/\psi (\to l^+l^-)\,\phi (\to K^+K^-)$ 
the explicit form of the distribution functions (\ref{angle_dist_1}) are given
by
\begin{eqnarray}
f_1(\mbox{cos}\Theta_{l^+},\chi;t) &=&
   ~~\frac{4}{3}b_1(t)\,\mbox{sin}^2\Theta_{l^+}
  +\frac{4}{3}b_2(t)\,(1-\mbox{sin}^2\Theta_{l^+}\mbox{cos}^2\chi)
\nonumber\\ &&
  +\frac{4}{3}b_3(t)\,(1-\mbox{sin}^2\Theta_{l^+}\mbox{sin}^2\chi)
  -{\cal O}_4(t)\,\mbox{sin}^2\Theta_{l^+}\mbox{sin} 2\chi\,,
\nonumber\\
f_2(\chi,t) &=&
   ~~b_1(t) +\frac{1}{2}b_2(t)\,(3-2\mbox{cos}^2\chi)
\nonumber\\ &&
  +\frac{1}{2}b_3(t)\,(3-2\mbox{sin}^2\chi)
  -b_4(t)\,\mbox{sin} 2\chi\,.
\label{f_12}
\end{eqnarray}

  The procedure of random generation of the variables $t$, $\chi$, 
$\mbox{cos}\,\Theta_{l^+}$ and $\mbox{cos}\,\Theta_{K^+}$ is as follows:
\begin{itemize}
\item First, the proper time $t$ is randomly generated according to the 
      distribution function $f_3(t)$.
\item Second, the angle $\chi$ is randomly generated according to 
      the single-dimensional distribution given by function $f_2(\chi;t)$ 
      with $t$ being fixed to be equal to the time, generated at the first
      step.
\item Then, the value $\mbox{cos}\,\Theta_{l^+}$ is generated according to
      distribution function $f_1(\Theta_{l^+},\chi;t)$ with values of $t$ 
      and $\chi$ fixed to be equal to their values generated at the 
      previous two steps.
\item Finally, the value of $\mbox{cos}\,\Theta_{K^+}$ is generated 
      according to the single-dimensional distribution given by function 
      $f_0(\Theta_{l^+},\Theta_{K^+},\chi;t)$ with a properly fixed values 
      of $t$, $\chi$ and $\mbox{cos}\,\Theta_{l^+}$.
\end{itemize}

   Two Monte Carlo methods of random numbers generation are implemented in 
the C++ class {\tt T\_VertexB\_VllVpp} of the {\tt BB\_dec} program of the 
SIMUB package \cite{SIMUB}.

   The first method is based on filling of the single-dimensional array 
{\tt f4\_Integ[n4\_cells]} of real numbers which represents numerically the 
four-dimensional distribution function $f_0(\Theta_{l^+},\Theta_{K^+},\chi;t)$.
   The array {\tt f4\_Integ} contains 
${\tt n4\_cells}= N(\mbox{cos}\Theta_{l^+})\times N(\mbox{cos}\Theta_{K^+})
                  \times N(\chi)\times N(t)$
elements, where $N(\mbox{cos}\Theta_{l^+})$, $N(\mbox{cos}\Theta_{K^+})$, 
$N(\chi)$ and $N(t)$ are the numbers of bins (generator resolutions) for 
corresponding variables randomly generated in a four-dimensional volume 
$V\in\{\mbox{cos}\Theta_{l^+},\mbox{cos}\Theta_{K^+},\chi,t\}$.
   A fast generation of these variables is performed by the function 
{\tt GetRandom4} by using the algorithm which is analogous to that was 
realized in the ROOT class {\tt TF3} \cite{ROOT} for generation of three 
random numbers distributed according to a three-dimensional probability 
function.
   In this case the array {\tt f4\_Integ} is filled in constructor of the 
class {\tt T\_VertexB\_VllVpp} where a large memory space for the array is 
reserved during all time of existence of this class sample.

   The second method involves sequential generation of random numbers 
according to approach based on usage of the single-dimensional distribution 
functions given by Eqs.~(\ref{time_dist}) and (\ref{f_12}).
   The corresponding algorithm of generation of the variables $t$, $\chi$, 
$\mbox{cos}\,\Theta_{l^+}$ and $\mbox{cos}\,\Theta_{K^+}$ is described above.
   In this case the constructor {\tt T\_VertexB\_VllVpp} does not fill any
arrays and, therefore, it does not reserve a large memory space.

   Both methods demonstrated the identical results with reasonable CPU time 
and memory used.

   The $B$-decay dynamics described by angular correlations (\ref{angle_dist}) 
should be included without fail into Monte Carlo generators developed for 
physics studies.
   The main motivation for developing the new dedicated package SIMUB was 
that already existing generators do not take into account all theoretical 
refinements which are of a great importance for Monte Carlo studies of 
$B$-decay dynamics.
   In particular, in the generators PYTHIA~\cite{PYTHIA} and QQ~\cite{QQ}
the time-dependent spin-angular correlations between the final-state particles
in the decays
$B^0_s(t),\overline{B}^0_s(t)\to J/\psi (\to l^+l^-)\,\phi (\to K^+K^-)$ 
are not reproduced in a proper way\footnote{In the latest version of the 
generator EvtGen~\cite{EvtGen} this mode can be also simulated including the
full time-dependent spin-angular correlations.}.

   The time and angular distributions in the helicity frame for decay 
$B^0_s(t),\overline{B}^0_s(t)\to J/\psi(\to\mu^+\mu^-)\,\phi(\to K^+K^-)$, 
generated by SIMUB, PYTHIA and QQ packages, are compared in 
Fig.~3.
   In case of PYTHIA usage, Fig.~3 shows the uniform distributions
for angular variables $\mbox{cos}\,\Theta_{K^+}$, $\mbox{cos}\,\Theta_{l^+}$ 
and $\chi$ because of lack of time-dependent angular correlations.
   Due to this simple reason PYTHIA can not be used for Monte Carlo studies of
dynamics of sequential two body decays of $B$ mesons in the channels of the
type $B\to V_1(\to\mu^+\mu^-)\, V_2(\to P^+P^-)$ with intermediate vector 
mesons $V_1$ and $V_2$.
   Unfortunately, another well known package QQ turns out to be also not 
suitable for study of decays of this type because of lack of azimuthal-angle 
$\chi$ correlations.

%------------------------------
 \section{Monte Carlo studies}
%------------------------------
    For Monte Carlo studies of the estimation of physical parameters by 
applying the angular-moments method, untagged samples of events of the decays
$B^0_s(t),\overline{B}^0_s(t)\to J/\psi\,\phi$ have been generated by using 
the package SIMUB with various sets of the input values of initial amplitudes
$|A_0(0)|$ and $|A_\bot (0)|$ and $\Delta\Gamma_s$.
    Other parameters are fixed as follows:
$$
\delta_1 =\pi\,,\,\, \delta_2 =0\,,
\Gamma_s = 1/\tau_s= 2.278~[\mbox{mm/c}]^{-1}\,,\,\,
\phi_c^{(s)}=0.04\,.
$$
   The value of $\Gamma_s$ used corresponds to the lifetime $\tau_s=1.464$ ps 
\cite{PDG} while the CP-violating weak phase $\phi_c^{(s)}$ was fixed as 
the upper limit of the constrain (\ref{weak-phase_ccs}).
   The value of $\Delta\Gamma_s$ is expected to be negative in the SM.
   The combined experimental result for $|\Delta\Gamma_s|/\Gamma_s$ 
is not precise: $|\Delta\Gamma_s|/\Gamma_s < 0.52$ at 95\% CL~\cite{PDG}.
   In the approximation of the equal $B^0_s$ and $B^0_d$ lifetimes, the 
$|\Delta\Gamma_s|$ extraction can be improved \cite{PDG}:  
$|\Delta\Gamma_s|/\Gamma_s < 0.31$ at 95\% CL.
   A set of the untagged-event samples has been generated with 
$\Delta\Gamma_s/\Gamma_s\in [-0.3,-0.01]$ to study the influence of
$\Delta\Gamma_s$ value on the estimation of $B^0_s(t)\to J/\psi\,\phi$ 
decay parameters from data analysis.

   The values of the time integrated observables $\tilde{b}_i^{(exp)}(T_0)$,
defined by Eq.~(\ref{observ_tild}), can be extracted from data according to 
Eq.~(\ref{observ_tilde_exp}) by summation of weighting functions for each 
event.
   The statistical error of $\tilde{b}_i(T_0)$ is defined by
\begin{equation}
(\delta \tilde{b}_i)^{(stat)} = \frac{1}{N(T)}\sqrt{\sum_{j=1}^{N(T_0)}
                                (\tilde{b}_i^{(exp)} - w_i^j)^2}\,,
\label{stat_err}
\end{equation}
while a systematic error due to limited precision of angular measurements can
be estimated as
\begin{equation}
(\delta \tilde{b}_i)^{(sys)} = 
\sqrt{\frac{\sum_{j=1}^{N(T_0)} \Delta_i^j}{N(T)}}\,.
\label{sys_err}
\end{equation}
  Here
$$
\Delta_i^j =  \bigg[\frac{\partial w_i^j}{\partial \cos{\Theta_{l^+}}}
                    \Delta (\cos{\Theta_{l^+}}) \bigg]^2
             +\bigg[\frac{\partial w_i^j}{\partial \cos{\Theta_{K^+}}}
                    \Delta (\cos{\Theta_{K^+}}) \bigg]^2 
             +\bigg[\frac{\partial w_i^j}{\partial \chi}
                    \Delta (\chi) \bigg]^2\,.
$$

   In a similar way, the values of the observables $\hat{b}_i^{(exp)}(T_0)$, 
defined by Eq.~(\ref{observ_hat}), can be extracted from the data according to 
Eq.~(\ref{observ_hat_exp}).
   The formulae for statistical and systematic errors for
$\hat{b}_i^{(exp)}(T_0)$ can be obtained by replacement of $w^j_i$ to $W^j_i$
in Eq.~(\ref{stat_err}) and the following redefinition of $\Delta^j_i$ in 
Eq.~(\ref{sys_err}):
$$
\Delta_i^j =  \bigg[\frac{\partial W_i^j}{\partial \cos{\Theta_{l^+}}}
                    \Delta (\cos{\Theta_{l^+}}) \bigg]^2
             +\bigg[\frac{\partial W_i^j}{\partial \cos{\Theta_{K^+}}}
                    \Delta (\cos{\Theta_{K^+}}) \bigg]^2 
             +\bigg[\frac{\partial W_i^j}{\partial \chi}
                    \Delta (\chi) \bigg]^2
             +\bigg[\frac{\partial W_i^j}{\partial t}
                    \Delta (t) \bigg]^2\,.
$$

   Eq.~(\ref{sys_err}) can be applied to estimate the systematic errors 
related both to the measurement precision of the detector and to the limited 
resolution of the Monte Carlo generator.
   In the SIMUB generator, for each variable
$V\in\{\mbox{cos}\Theta_{l^+},\mbox{cos}\Theta_{K^+},\chi,t\}$ randomly
generated for decays $B^0_s(t)\to J/\psi (\to l^+l^-)\,\phi (\to K^+K^-)$, the 
number of bins in the region $[V_{min},V_{max}]$ was set as $N^{gen}=50\,000$.
   The generation precision for the variable $V$ is defined as 
$\Delta (V)=(V_{max}-V_{min})/N^{gen}$
and systematic errors~(\ref{sys_err}) are proportional to $(N^{gen})^{-1/2}$.
   The $B$-meson proper time was generated within the interval
$t\in [0, T=2\,\mbox{mm/c}]$ which includes 99.3\% of all $B$-decays.
   We have used samples with a maximum of 100\,000 events of the decay 
$B_s^0\to J/\psi\,\phi$ because a statistics of about 80\,000 events is 
expected to be obtained per year at the CMS detector at the LHC low luminosity
under realistic triggering conditions \cite{stepanov}.

   Table~\ref{tab:MC_Tilde_b_A_B} shows the values of the observables
$\tilde{b}_i^{(exp)}(T)\equiv\tilde{b}_i^{(exp)}(T_0=T)$ extracted from the 
Monte Carlo data by applying the sets A and B of weighting functions, given 
by Eqs.~(\ref{set_A}) and (\ref{set_B}), respectively.
   Various theoretical models for estimation of the transversity 
amplitudes $|A_0(0)|$ and $|A_\bot (0)|$  (see Table \ref{tab:observ_theor})  
have been considered to fix these parameters in the SIMUB generator.
   It can be seen from Table~\ref{tab:MC_Tilde_b_A_B} that the choice of 
$N^{gen}=50\,000$ provides negligibly small systematic errors for the 
observables as compared with the statistical ones.
   Moreover, both errors slightly depend on the values of the observables.
   For observables obtained by using the set-B weighting functions, the 
statistical errors are significantly smaller than in case of the set-A
weighting functions.
   We should also note that even with the statistics of $100\,000$ events, the
values of observables $\tilde{b}^{(exp)}_{4,6}(T)$ and -- as consequence of
Eq.~(\ref{sin_phase}) -- the combination 
$\mbox{cos}\delta_{1,2}\,\mbox{sin}\phi$ cannot be extracted from the data if
the CP-violating weak phase $\phi_c^{(s)}$ is small according to the SM 
expectation (\ref{weak-phase_ccs}). 
   In this case, these parameters can be estimated only by using a 
statistics which is not less than $3\times 10^9$ $B^0_s(t)\to J/\psi\,\phi$ 
decays.

   Analysis of the same Monte Carlo data leads to similar conclusions 
concerning the behavior of statistical and systematic errors for the 
observables $\hat{b}_i^{(exp)}(T)\equiv\hat{b}_i^{(exp)}(T_0=T)$.
   To illustrate the performance of our method in this case, only the results 
obtained for transversity amplitudes, corresponding to the Cheng's model
\cite{cheng}, are shown in Table~\ref{tab:MC_Hat_b_A_B}.

   Fig.~4 shows the dependence of the observables $\tilde{b}_i(T)$ 
and 
$$
\hat{b}^{\,\prime}_i(T) \equiv 
\frac{1-\mbox{e}^{-\Gamma_sT}}{\Gamma_sT}\hat{b}_i(T)\quad (i=1,2,3)
$$ 
on the value of the ratio $\Delta\Gamma_s/\Gamma_s$.
   For $\Delta\Gamma_s = 0$ we have 
$$
\tilde{b}_{1,2,3}(T)|_{\Delta\Gamma_s=0}= 
\hat{b}^{\,\prime}_{1,2,3}(T)|_{\Delta\Gamma_s=0} = |A_{0,||,\bot}(0)|^2\,.
$$
   The observables $\tilde{b}_i(T)$ slightly depend on $\Delta\Gamma_s$.
   The rather strong dependence of the observables $\hat{b}_i(T)$ on the decay
width difference $\Delta\Gamma_s$, shown in Fig.~4, can be used for
extraction of this parameter from the data analysis as it will be discussed 
below.

   Under the assumption $\phi^{(s)}_c=0$, we have from Eq.~(\ref{G_hat}):
$$
\hat{G}^{(0)}_{L/H}(T)= 
\pm 2\frac{\mbox{e}^{\pm\Delta\Gamma_{L/H}T/2}-1}{\Delta\Gamma_{L/H}}\,.
$$
   Therefore, the values of the auxiliary parameters $\Delta \Gamma_{L/H}$, 
defined by Eq.~(\ref{Delta_G_L/H}), can be determined separately by using the
ratios of observables $\hat{b}_i^{(exp)}(T)/\hat{b}_i^{(exp)}(T_0)$, extracted
from the data analysis, and solving numerically the equations which arise from
one of the following relations:
\begin{eqnarray}
\hat{b}_i(T)/\hat{b}_i(T_0) = \hat{G}_L(T)/\hat{G}_L(T_0)
\qquad (i=1,2,5)
\label{hbT-d-hbT0}
\end{eqnarray}
-- to determine $\Delta\Gamma_L$, and the relation
\begin{eqnarray}
\hat{b}_3(T)/\hat{b}_3(T_0) = \hat{G}_H(T)/\hat{G}_H(T_0)
\label{hb3T-d-hb3T0}
\end{eqnarray}
-- to determine $\Delta\Gamma_H$.
   Then, the decay-width parameters $\Gamma_s$, $\Delta\Gamma_s$ and 
$\Gamma_{L/H}$ can be determined via $\Gamma^\prime$ and $\Delta\Gamma_{L/H}$
as
\begin{equation}
\Gamma_s =\Gamma^{\prime} - \frac{\Delta\Gamma_L - \Delta\Gamma_H}{4}\,,\qquad
\Delta\Gamma_s = \frac{\Delta\Gamma_L + \Delta\Gamma_H}{2}\,,\qquad
\Gamma_{L/H} = \Gamma^{\prime} \mp \frac{\Delta\Gamma_L}{2}\,.
\label{G_L_H-phys}
\end{equation}
   So, by using some reasonable approximation for $\Gamma^\prime$ as a starting
point for the data analysis, the experimental value of $\Gamma_s$ can be 
essentially improved simultaneously with determination of $\Delta\Gamma_s$.
   The statistical error of $\Gamma_s$ determination is expected to be
twice smaller than for $\Delta\Gamma_s$ determination.

   The direct numerical calculations have shown that the difference between the
values of observables $\hat{b}_i(T)$ $(i=1,2,3,5)$, calculated with 
$\phi^{(s)}_c=0$ and $\phi^{(s)}_c=0.04$, does not exceed 0.01\%.
   Even in case of statistics of 100~000 events this difference is negligibly
small as compared with statistical errors for these observables (see 
Table~\ref{tab:MC_Hat_b_A_B}).
   Therefore, the assumption $\phi^{(s)}_c=0$ is a good approximation for
$\Gamma_s$, $\Delta\Gamma_s$ and $\Gamma_{L/H}$ determination by the method 
described above.

    Table~\ref{tab:DG_L_H_Measurement} shows the results of determination of 
the decay-width parameters after applying the described procedure to the Monte
Carlo data.
   The sample of 100\,000 events generated in case of Cheng's model with 
$\Delta\Gamma_s/\Gamma_s = -0.15$ has been used.
   Both sets A and B of weighting functions have been applied to extract 
the observables $\hat{b}_i^{(exp)}$.
   The value of $\Gamma^\prime$, which is treated as some arbitrary initial
approximation for the total decay width of $B^0_s$-meson, was fixed as 
$\Gamma^{\prime} = 1.05\,\Gamma_s$, i.e. it was shifted by 5\% relative to
the ``true'' value of $\Gamma_s$ fixed in the Monte Carlo generator SIMUB.
   The value of $T_0 = 0.1\,T$ was chosen as it provides the minimal 
statistical errors to determine the ratios 
$\hat{b}_i^{(exp)}(T)/\hat{b}_i^{(exp)}(T_0)$.
   In Table~\ref{tab:DG_L_H_Measurement} we present the result for 
$\Delta\Gamma_L$ obtained from the ratio 
$\hat{b}_1^{(exp)}(T)/\hat{b}_1^{(exp)}(T_0)$ only, which gives the best 
precision.
   Table~\ref {tab:DG_L_H_Measurement} shows that the set-B weighting 
functions give more precise and stable results than the set-A functions.

   To improve the precision of $\Delta\Gamma_s$ determination, the same 
procedure should be repeated with $\Gamma^\prime$ fixed to be equal to the 
value of $\Gamma_s$ determined at the first step.
   Because of $\Delta\Gamma_s = \Delta\Gamma_L =\Delta\Gamma_H$ in case of
$\Gamma^\prime = \Gamma_s$, the value of $\Delta\Gamma_s$ is defined at 
the second step to be equal to the value of $\Delta\Gamma_L$ determined 
from the ratio $\hat{b}_1^{(exp)}(T)/\hat{b}_1^{(exp)}(T_0)$ using 
Eq.~(\ref{hbT-d-hbT0}). 
   Using the values of $\Delta\Gamma_s$ from 
Table~\ref{tab:DG_L_H_Measurement} as an input value of $\Gamma^\prime$ at the 
second step, we have obtained finally the following results (to be compared 
with the input value $\Delta\Gamma_s = -0.3418$ set in the SIMUB generator): 
\begin{eqnarray*}
\Delta\Gamma_s^{exp} &=& -0.330\pm 0.057\pm 0.004 \qquad (\mbox{set~A})\,,
\\
\Delta\Gamma_s^{exp} &=& -0.338\pm 0.034\pm 0.002 \qquad (\mbox{set~B})\,.
\end{eqnarray*}
   This way one can reduce not only the statistical error but also 
essentially improve the stability of the $\Delta\Gamma_s$ result even in case
of using the set-A weighting functions.

   Table~\ref{tab:DG_Measurement_NEv} shows the statistical errors of 
$\Delta\Gamma_s/\Gamma_s$ determination by the described approach applied 
to different statistics of Monte Carlo events generated with various 
"true" values of $\Delta\Gamma_s$.
   The lack of numbers in the table corresponds to cases when the approach
is not able to give a certain result for $\Delta\Gamma_s$.
   The use of set-B weighting functions gives more stable results even in 
case of too small statistics and values of $\Delta\Gamma_s$, for which the 
same approach does not work with set-A functions.
   The sensitivity of the method is measured by the statistical error of
$\Delta\Gamma_s/\Gamma_s$, which only slightly depends on the 
value of this ratio and is proportional to $1/\sqrt{N}$, where $N$ is the 
number of events.
   In particular, for a statistics 100\,000 events, the statistical error is 
about 0.015, while for 1000 events -- about 0.15.

   The value of $\Delta\Gamma_s$ can be determined similarly by 
using the ratios $\tilde{b}_i^{(exp)}(T)/\tilde{b}_i^{(exp)}(T_0)$ or 
$\hat{b}_i^{(exp)}(T)/\tilde{b}_i^{(exp)}(T)$, extracted from the data analysis
with $\Gamma^\prime=\Gamma_s$, and solving the equations arising from the 
relations
$$
\tilde{b}_i(T)/\tilde{b}_i(T_0) =
\tilde{G}_L(T)/\tilde{G}_L(T_0)\quad (i=1,2,5)\,,\qquad
\tilde{b}_3(T)/\tilde{b}_3(T_0) = \tilde{G}_H(T)/\tilde{G}_H(T_0) 
$$
or
$$
\hat{b}_i(T)/\tilde{b}_i(T) =
\hat{G}_L(T)/\tilde{G}_L(T)\quad (i=1,2,5)\,,\qquad
\hat{b}_3(T)/\tilde{b}_3(T) = \hat{G}_H(T)/\tilde{G}_H(T)\,.
$$
   But in both these cases the precision of $\Delta\Gamma_s$ determination
turns out to be worse than in the approach based on the ratios 
$\hat{b}_i^{(exp)}(T)/\hat{b}_i^{(exp)}(T_0)$ because of the weak 
$\Delta\Gamma_s$-dependence of the $\tilde{b}_i(T)$ observables. 

   The initial transversity amplitudes and strong-phase difference can be
recalculated from the values of observables $\tilde{b}^{(exp)}_i(T)$ according
to Eq.~(\ref{observ_tild_ut_3}).
   The results of such determination of the parameters 
$|A_f(0)|^2$ $(f=0,||,\bot)$ and $\mbox{cos}(\delta_2 - \delta_1)$
are shown in Table~\ref{tab:tilde_b_Measurement} for different statistics.
   We have used the Monte Carlo sample generated with the theoretical values 
of the amplitudes $|A_0(0)|$ and $|A_\bot (0)|$ corresponding to Cheng's model
\cite{cheng}.
   To extract the observables $\tilde{b}^{(exp)}_i(T)$, the set B of the 
weighting function has been applied to Monte Carlo data. 
   To estimate the statistical errors for parameters $|A_f(0)|^2$ 
$(f=0,||,\bot)$ and $\mbox{cos}(\delta_2 - \delta_1)$, the standard 
error-propagation method has been applied to the statistical errors of the 
observables $\tilde{b}^{(exp)}_i(T)$, taking into account the correlation
between pairs of different observables.
   The systematic errors of the observables related to the limited generator 
resolution are neglected.
   The total errors for parameters $|A_f(0)|^2$ $(f=0,||,\bot)$ should also
include the additional uncertainty related to the error of calculation of
$\tilde{\gamma}(T)$ caused by the error of $\Delta \Gamma_s$ (see definition 
of $\tilde{\gamma}(T)$ in Eq.~(\ref{gamma}) and Eq.~(\ref{G_tilde})).
   In Table~\ref{tab:tilde_b_Measurement} we also show these errors calculated 
by assuming $\Delta\Gamma_s = -0.15\,\Gamma_s$ 
(see Table~\ref{tab:DG_Measurement_NEv})
\begin{eqnarray}
\frac{\delta(\Delta\Gamma_s)}{|\Delta\Gamma_s|}
 = \left\{
     \begin{array}{lr}
       30 \%   \qquad\mbox{for 10\,000 events},  \\
       9.3\%  \qquad\mbox{for 100\,000 events}. \\
     \end{array}
   \right.
\label{DG_err_esimat}
\end{eqnarray}

%-------------------
\section{Conclusion}
%-------------------

   For the decay $B^0_s\to J/\psi\,\phi$ in the framework of the method of 
angular moments a non-fit scheme for separate estimation of the parameters
$\Delta\Gamma_s$, $\Gamma_s$ and $|A_f(0)|^2$ $(f=0,||,\bot)$ has been 
proposed, based on analysis of an untagged sample, and studied by Monte Carlo
method.
   A strong dependence of statistical measurement errors on the choice of the
weighting functions has been demonstrated.
   The statistical error of the ratio $\Delta\Gamma_s/\Gamma_s$ for values in 
the interval [0.03, 0.3] was found to be independent of the central value and
amounts to about 0.015 for $10^5$ events. 
   The method of angular moments gives stable results for the estimate of
$\Delta\Gamma_s$ and is found to be an efficient and flexible tool for the
quantitative investigation of the $B^0_s\to J/\psi\,\phi$ decay.

%--------------------------
\section*{Acknowledgements}
%--------------------------

   We would like to thank G.~Bohm, J.~Mnich for the careful reading of the 
manuscript and useful discussions.
\newpage

\appendix
%=================================================
\section{General information on the SIMUB package}
%=================================================

   The generator SIMUB was developed for Monte Carlo simulation of $B$-meson 
production and decays at the CMS detector.
   The SIMUB package provides two regimes of $B$-decay generation:
\begin{itemize}
\item simulation of decay after fragmentation according to angular 
      distribution and time dependence (in case of a neutral $B$-meson) 
      governed by spin-angular correlations, time evolution and other 
      theoretical refinements -- {\it dynamical mode},
\item simulation of decay by PYTHIA into phase space -- {\it PYTHIA 
      kinematical  mode}.
\end{itemize}
   The package is kept under the directory {\tt SIMUB} which has the following
structure:
\begin{itemize}
\item {\tt bb\_gen} -- routines needed to generate events with $b\bar{b}$-pairs
                       at parton level (FORTRAN, PYTHIA, HBOOK);
\item {\tt bb\_frg} -- routines performing string fragmentation, generation
                       of $B$-mesons, and decays of particles in the PYTHIA 
                       kinematical mode (FORTRAN, PYTHIA, HBOOK); storing the 
                       results into standard HEPEVT or PYJETS Ntuples for 
                       further usage in the CMS simulation or for analysis;
\item {\tt BB\_dec} -- routines performing $B$-decays in dynamical mode (C++, 
                       ROOT); 
\item {\tt include} -- a collection of common blocks for the routines
                       {\tt bb\_gen}, {\tt bb\_frg} and {\tt BB\_dec};
\item {\tt lib} -- the PYTHIA source codes and object files;
\item {\tt doc} -- documentation.
\end{itemize}
   The data flow between the three main parts ({\tt bb\_gen}, {\tt bb\_frg},
and {\tt BB\_dec}) of the package SIMUB is shown in Fig.~5 and organized as 
follows:
\begin{itemize}
\item At the first stage the events containing $b\bar{b}$-pairs are generated 
      by the program {\tt bb\_gen}, stopped before fragmentation and written 
      to an intermediate Ntuple to store events at parton level.
\item At the next step the program {\tt bb\_frg} reads $b\bar{b}$-events from
      the intermediate Ntuple and performs the string fragmentation and the
      decays of all particles in the PYTHIA kinematical mode with exception of
      $B$-mesons selected according to user directives to be decayed in 
      dynamical mode.
\item Then, the information about the selected $B$-mesons is used by the 
      program {\tt BB\_dec} to perform their decays according to dynamical 
      modes.
\item Finally, the information about the selected $B$-mesons and products of
      their decays is transfered to the subprogram {\tt bb\_frg} to be added
      to the information about other particles and stored in a final output 
      Ntuple.
\end{itemize}

   The package SIMUB is installed and tested on Linux (RedHat 6.x, 7.x) 
platforms.
   The package and documentation for user is found on the SIMUB Package Home 
Page:
\begin{center}
http://cmsdoc.cern.ch/\~\,shulga/SIMUB/SIMUB.html.
\end{center}
   To install the program, the file {\tt SIMUB.tar.gz} should be copied
from the SIMUB Package Home Page to the user directory and unpacked 
to obtain the directory {\tt SIMUB}.

   The current version of the program SIMUB is adopted to the usage of the
generator PYTHIA \cite{PYTHIA} (version 6.215), CERNLIB-2002 \cite{CERN_lib}, 
and ROOT package \cite{ROOT} (version 3.05/07).
   The standard CERN Program Library should be installed at a Linux machine 
according to the record in the files {\tt bb\_gen/mak/Makefile} and 
{\tt bb\_frg/mak/Makefile}:
\begin{verbatim}
CERNLIB := `cernlib pawlib graflib mathlib packlib kernlib`
\end{verbatim}
or the record in the file {\tt BB\_dec/mak/Makefile}:
\begin{verbatim}
SYSLIBS = -L/cer/new/lib/ -lpacklib ...
\end{verbatim}
   The user has to generate the object file of the PYTHIA generator in the 
directory {\tt lib} and to install the ROOT package.
   Detailed instructions on the installation, compilation and running
of the package SIMUB are given in the documents ``User and Developer Guide'' 
and ``Quick Start''.
  The postscript files of these documents are placed in the directory
{\tt SIMUB/doc} and also found on the SIMUB Package Home Page.
 
\newpage

%===========================================================================
%                               References
%===========================================================================

\newpage

%=============================================================================
%                               Table 1
%=============================================================================
\begin{table}
\caption{Predictions for $B^0_s\to J/\psi\,\phi$ (in brackets -- for
         $B^0_d\to J/\psi\,K^\star$) observables obtained in Ref.~\cite{dighe2}
         for various model estimates of the $B\to K^\star$ form factors 
         \cite{wirbel,soares,cheng} (the normalization condition
         $|A_0(0)|^2+|A_{||}(0)|^2+|A_\bot (0)|^2=1$ is implied)}
\label{tab:observ_theor}
%-----------------------------------------------------------------------------
\vspace*{3mm}
\begin{center}
\begin{tabular}{|c|c|c|c|}
\hline
Observable & BSW \cite{wirbel} & Soares \cite{soares} & Cheng \cite{cheng}\\
\hline
\hline
$|A_0(0)|^2$     & 0.55 (0.57) & 0.41 (0.42) & 0.54 (0.56) \\ \hline
$|A_\bot (0)|^2$ & 0.09 (0.09) & 0.32 (0.33) & 0.16 (0.16) \\ \hline
\end{tabular}
\end{center}
\end{table}
\vspace*{100mm}
%=============================================================================

%=============================================================================
%                             Table 2
%=============================================================================
\begin{table}
\begin{center}
\caption{Comparison of the observables $\tilde{b}^{(exp)}_i(T)$, extracted 
         from Monte Carlo data, with their values $\tilde{b}^{(th)}_i(T)$ 
         corresponding to various theoretical models for $|A_0(0)|$ and 
         $|A_\bot (0)|$. A sample of 100\,000 decay events generated with
         $\Delta\Gamma_s/\Gamma_s=-0.15$ was used. The first errors are 
         statistical (see Eq.~(\ref{stat_err})) while the second errors 
         correspond to the systematic uncertainties only from limited 
         angular precision (see Eq.~(\ref{sys_err}))}
\vspace*{3mm}
\label{tab:MC_Tilde_b_A_B}
%----------------------------------------------------------------------------
\begin{center}
a) BSW model \cite{wirbel}:
\end{center}
\begin{tabular}{|c|c|c|c|}
\hline
$i$& $\tilde{b}_i^{(th)}(T)$ & $\tilde{b}^{(exp)}_i(T)$ (set A)
                             & $\tilde{b}^{(exp)}_i(T)$ (set B)\\
\hline
\hline
1& 0.5425  & $ 0.5409\pm 0.0044\pm 0.0003$ & $ 0.5432\pm 0.0024\pm 0.0002$ \\[-2mm]
2& 0.3551  & $ 0.3619\pm 0.0047\pm 0.0004$ & $ 0.3579\pm 0.0036\pm 0.0004$ \\[-2mm]
3& 0.1024  & $ 0.0972\pm 0.0049\pm 0.0004$ & $ 0.0991\pm 0.0034\pm 0.0004$ \\[-2mm]
4&-0.00055 & $-0.0017\pm 0.0037\pm 0.0004$ & $-0.0021\pm 0.0033\pm 0.0003$ \\[-2mm]
5&-0.4389  & $-0.4344\pm 0.0050\pm 0.0003$ & $-0.4344\pm 0.0050\pm 0.0003$ \\[-2mm]
6& 0.00067 & $ 0.0037\pm 0.0055\pm 0.0003$ & $ 0.0037\pm 0.0055\pm 0.0003$ \\
\hline
\end{tabular}
%----------------------------------------------------------------------------
\begin{center}
b)  Model by Soares \cite{soares}:
\end{center}
\begin{tabular}{|c|c|c|c|}
\hline
$i$& $\tilde{b}_i^{(th)}(T)$ & $\tilde{b}^{(exp)}_i(T)$ (set A)
                             & $\tilde{b}^{(exp)}_i(T)$ (set B)\\
\hline
\hline
1& 0.3908  & $ 0.3900\pm 0.0046\pm 0.0003 $ & $ 0.3955\pm 0.0023\pm 0.0002 $ \\[-2mm]
2& 0.2574  & $ 0.2617\pm 0.0049\pm 0.0004 $ & $ 0.2551\pm 0.0037\pm 0.0004 $ \\[-2mm]
3& 0.3518  & $ 0.3483\pm 0.0047\pm 0.0004 $ & $ 0.3509\pm 0.0037\pm 0.0004 $ \\[-2mm]
4&-0.00086 & $-0.0083\pm 0.0040\pm 0.0004 $ & $-0.0017\pm 0.0035\pm 0.0003 $ \\[-2mm]
5&-0.3171  & $-0.3156\pm 0.0052\pm 0.0003 $ & $-0.3156\pm 0.0052\pm 0.0003 $ \\[-2mm]
6& 0.0011  & $ 0.0008\pm 0.0052\pm 0.0003 $ & $ 0.0008\pm 0.0052\pm 0.0003 $ \\\hline
\end{tabular}
%----------------------------------------------------------------------------
\begin{center}
c) Model by Cheng \cite{cheng}:
\end{center}
\begin{tabular}{|c|c|c|c|}
\hline
$i$& $\tilde{b}_i^{(th)}(T) $ & $\tilde{b}^{(exp)}_i(T)$ (set A)
                              & $\tilde{b}^{(exp)}_i(T)$ (set B)\\ \hline\hline
1& 0.5271  & $ 0.5228\pm 0.0045\pm 0.0003$ & $ 0.5267\pm 0.0024\pm 0.0001$ \\[-2mm]
2& 0.2928  & $ 0.2980\pm 0.0048\pm 0.0004$ & $ 0.2950\pm 0.0036\pm 0.0004$ \\[-2mm]
3& 0.1801  & $ 0.1791\pm 0.0048\pm 0.0004$ & $ 0.1778\pm 0.0035\pm 0.0004$ \\[-2mm]
4&-0.00066 & $-0.0030\pm 0.0037\pm 0.0003$ & $-0.0034\pm 0.0034\pm 0.0003$ \\[-2mm]
5&-0.3928  & $-0.3927\pm 0.0051\pm 0.0003$ & $-0.3927\pm 0.0051\pm 0.0003$ \\[-2mm]
6& 0.00088 & $-0.0019\pm 0.0054\pm 0.0003$ & $-0.0019\pm 0.0054\pm 0.0003$ \\  \hline
\end{tabular}
\end{center}
\end{table}
%=============================================================================

%=============================================================================
%                             Table 3
%=============================================================================
\begin{table}
\begin{center}
\caption{Comparison of the values of observables $\hat{b}^{(exp)}_i(T)$,
         with $\Gamma' = \Gamma_s$, extracted from the Monte Carlo data,
         with their values $\hat{b}^{(th)}_i(T)$ corresponding to the model of
         Cheng \cite{cheng} for initial transversity amplitudes }

\vspace*{3mm}
\label{tab:MC_Hat_b_A_B}
%----------------------------------------------------------------------------
\begin{tabular}{|c|c|c|c|}
\hline
$i$& $\hat{b}_i^{(th)}(T) $ & $\hat{b}^{(exp)}_i(T)$ (set A)
                            & $\hat{b}^{(exp)}_i(T)$ (set B)\\
\hline
\hline
1& 2.2036 & $ 2.176\pm 0.044\pm 0.003$ & $ 2.206\pm 0.026\pm 0.001$ \\
2& 1.2242 & $ 1.282\pm 0.045\pm 0.004$ & $ 1.245\pm 0.034\pm 0.004$ \\
3& 0.9187 & $ 0.917\pm 0.045\pm 0.004$ & $ 0.930\pm 0.034\pm 0.004$ \\
4&-0.0073 & $-0.099\pm 0.036\pm 0.003$ & $-0.094\pm 0.032\pm 0.003$ \\
5&-1.6425 & $-1.618\pm 0.048\pm 0.003$ & $-1.618\pm 0.048\pm 0.003$ \\
6& 0.0098 & $ 0.067\pm 0.050\pm 0.003$ & $ 0.067\pm 0.050\pm 0.003$ \\  \hline
\end{tabular}
\end{center}
\end{table}
%=============================================================================

%=============================================================================
%                             Table 4
%=============================================================================
\begin{table}
\begin{center}
\caption{Results of determination of the decay-width parameters (in units 
         (mm/$c$)$^{-1}$) based on extraction of the observables 
         $\hat{b}_i^{(exp)}$ from analysis of 100\,000 Monte Carlo events.
         The input value of $\Delta\Gamma_s$ corresponds to 
         $\Delta\Gamma_s/\Gamma_s = -0.15$ }
\label{tab:DG_L_H_Measurement}
\vspace*{-5mm}
\end{center}
%----------------------------------------------------------------------------
\begin{center}
\begin{tabular}{|c|c|c|c|}
\hline
Parameter        &Input value & Measurement (set A) 
                              & Measurement (set B) \\ \hline
$\Delta\Gamma_L$ & -0.1139    & $-0.103\pm 0.058\pm 0.003$
                              & $-0.110\pm 0.034\pm 0.002$ \\
$\Delta\Gamma_H$ & -0.5696    & $-0.478\pm 0.137\pm 0.012$
                              & $-0.554\pm 0.101\pm 0.012$ \\
$\Gamma_L$       &  2.4493    & $ 2.444\pm 0.029\pm 0.002$ 
                              & $ 2.447\pm 0.017\pm 0.001$ \\
$\Gamma_H$       &  2.1076    & $ 2.154\pm 0.068\pm 0.006$ 
                              & $ 2.115\pm 0.050\pm 0.006$  \\ 
$\Gamma_s$       &  2.2784    & $ 2.299\pm 0.037\pm 0.003$ 
                              & $ 2.281\pm 0.027\pm 0.003$ \\
$\Delta\Gamma_s$ & -0.3418    & $-0.290\pm 0.074\pm 0.006$
                              & $-0.332\pm 0.053\pm 0.006$ \\ \hline
\end{tabular}
\end{center}
\end{table}
%=============================================================================

%=============================================================================
%                                Table 5
%=============================================================================
\begin{table}
\begin{center}
\caption{Statistical errors of $ \Delta\Gamma_s$ extraction (in units
         $\mbox{[mm/c]}^{-1}$) obtained by applying the angular-moments 
         method with set-B (set-A) weighting functions to the Monte Carlo
         data samples with different numbers of events
        }
%\vspace*{3mm}
\label{tab:DG_Measurement_NEv}
\vspace*{-5mm}
\end{center}
%----------------------------------------------------------------------------
\begin{center}
\begin{tabular}{|c|c|c|c|c|c|}
\hline
  $\Delta\Gamma_s/\Gamma_s $  
              &200 events & 500 events &$10^3$ events&$10^4$ events& $10^5$ events  \\ \hline
-0.03&-        &-        &-          &0.035\,(-)    &0.014\,(0.023) \\
-0.05&-        &-        &-          &0.046\,(-)    &0.014\,(0.022) \\
-0.1 &-        &-        &0.11\,(-)   &0.046\,(0.079)&0.014\,(0.024) \\
-0.15&-        &-        &0.13\,(0.19)&0.045\,(0.078)&0.014\,(0.024) \\
-0.2 &-        &0.23\,(-)&0.12\,(0.18)&0.048\,(0.072)&0.015\,(0.026) \\
-0.3 &0.21\,(-)& 0.23\,(-)&0.18\,(0.20)&0.050\,(0.083)&0.016\,(0.028)\\ \hline
\end{tabular}
\end{center}
\end{table}
%=============================================================================

%=============================================================================
%                                Table 6
%=============================================================================
\begin{table}
\begin{center}
\caption{Determination of initial transversity amplitudes and strong-phase 
         difference by using the values of observables 
         $\tilde{b}^{(exp)}_i(T)$ extracted from Monte Carlo data.
         The events sample has been generated for the case of Cheng's model 
         \cite{cheng} for transversity amplitudes and with 
         $\Delta\Gamma_s = 0.15\,\Gamma_s$. The first errors are statistical 
         while the second errors are caused by uncertainties of 
         $\Delta\Gamma_s$ determination}
\label{tab:tilde_b_Measurement}
%----------------------------------------------------------------------------
\vspace*{5mm}
\begin{tabular}{|c|c|c|c|}
\hline
Parameter                      &Input value&10\,000 events &100\,000 events\\ 
\hline
$|A_0(0)|^2$                   & 0.54   &$ 0.527\pm 0.007 \pm 0.012 $
                                        &$ 0.5398\pm 0.0023\pm 0.0011$\\
$|A_{||}|^2$                   & 0.30   &$ 0.337\pm 0.011 \pm 0.008$
                                        &$ 0.3023\pm 0.0036\pm 0.0006$\\
$|A_\bot|^2$                   & 0.16   &$ 0.136 \pm 0.010 \pm 0.020$
                                        &$ 0.1579\pm 0.0032\pm 0.0018$\\
$\mbox{cos}(\delta_2-\delta_1)$&  -1    &$-1.021 \pm 0.044           $
                                        &$-0.9962\pm 0.015$\\
\hline
\end{tabular}
\end{center}
\end{table}
%=============================================================================
\newpage

\section*{Figure captions}

{\bf Figure 1.} Definition of physical angles for description of decays
                $B^0_s(t),\overline{B}^0_s(t)\to J/\psi (\to l^+l^-)\,\phi 
                (\to K^+K^-)$ in the helicity frame.

{\bf Figure 2.} Color suppressed diagrams for decays
                $B^0_q\to J/\psi V$ $((q,V)\in\{(s,\phi),(d,K^\star )\})$.

{\bf Figure 3.} Comparison of time and angular distributions for the decays 
                $B^0_s(t),\overline{B}^0_s(t)\to J/\psi(\to \mu^+\mu^-)\,
                \phi(\to K^+K^-)$ generated by SIMUB, PYTHIA and QQ packages.
                For time $t$ we use units 1mm/$c\approx 3.33\times 10^{-12}$ 
                sec.

{\bf Figure 4.} Dependence of the observables $\tilde{b}_i(T)$ and
                $\hat{b}^{\,\prime}_{i}(T)\equiv 
                \hat{b}_i(T)[1-\mbox{exp}(-\Gamma T)]/(\Gamma T)$ ($i=1,2,3$) 
                on the value of $\Delta\Gamma_s/\Gamma_s$. The observables 
                have been calculated for the case of Cheng's model for 
                transversity amplitudes.

{\bf Figure 5.} The flow of data within the package SIMUB.

\newpage
%=============================================================================
%                           Figure 1
%=============================================================================
\begin{center}
\epsfysize=0.55\textheight
\epsfbox{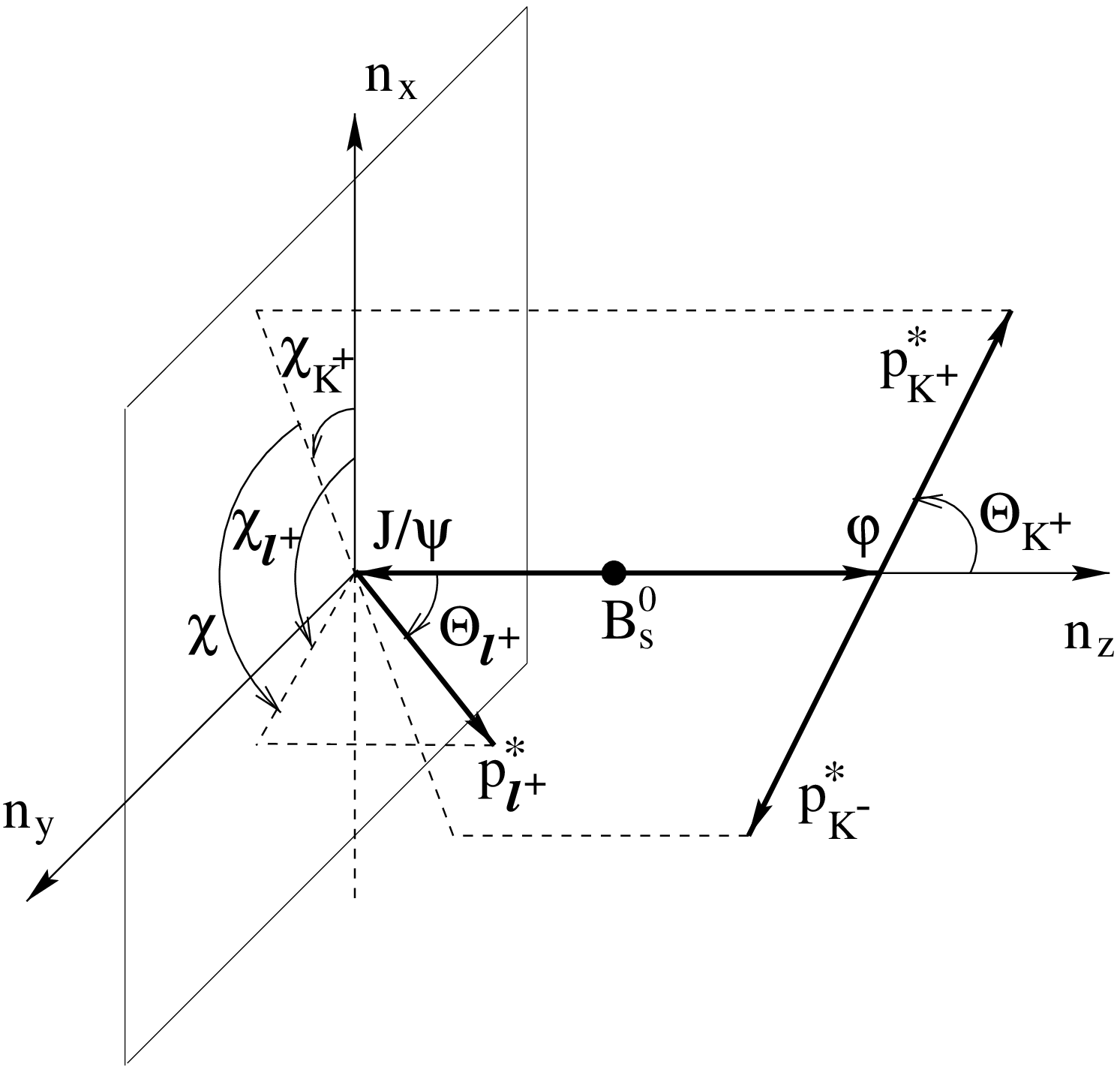}
\end{center}

\vspace{10mm}
{\bf Figure 1}
%=============================================================================

%=============================================================================
%                           Figure 2
%=============================================================================
\begin{center}
\epsfysize=0.25\textheight
\epsfbox{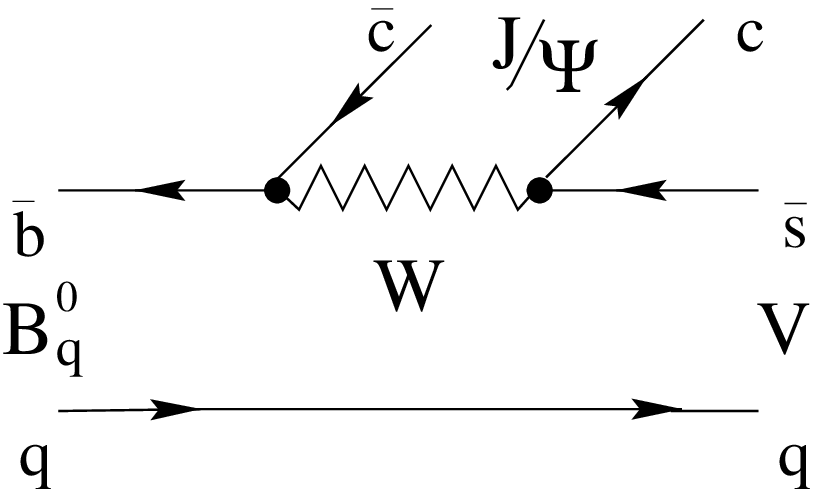}
\end{center}

\vspace{10mm}
{\bf Figure 2}

%=============================================================================

%=============================================================================
%                           Figure 3
%=============================================================================
\begin{center}
\epsfysize=0.65\textheight
\epsfbox{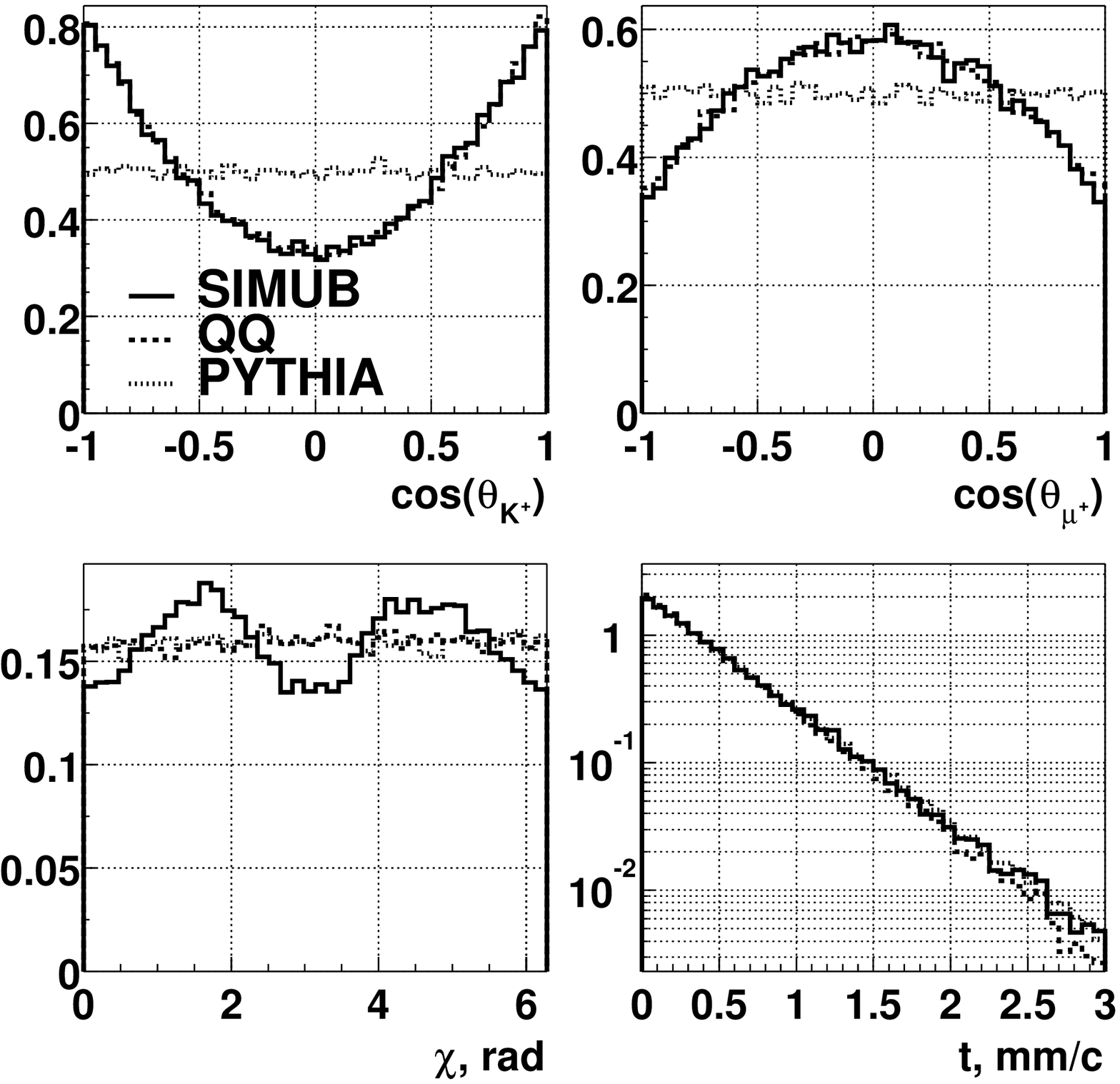}
\end{center}

\vspace{10mm}
{\bf Figure 3}
\newpage
%=============================================================================

%=============================================================================
%                           Figure 4
%=============================================================================
\begin{center}
\epsfysize=0.45\textheight
\epsfbox{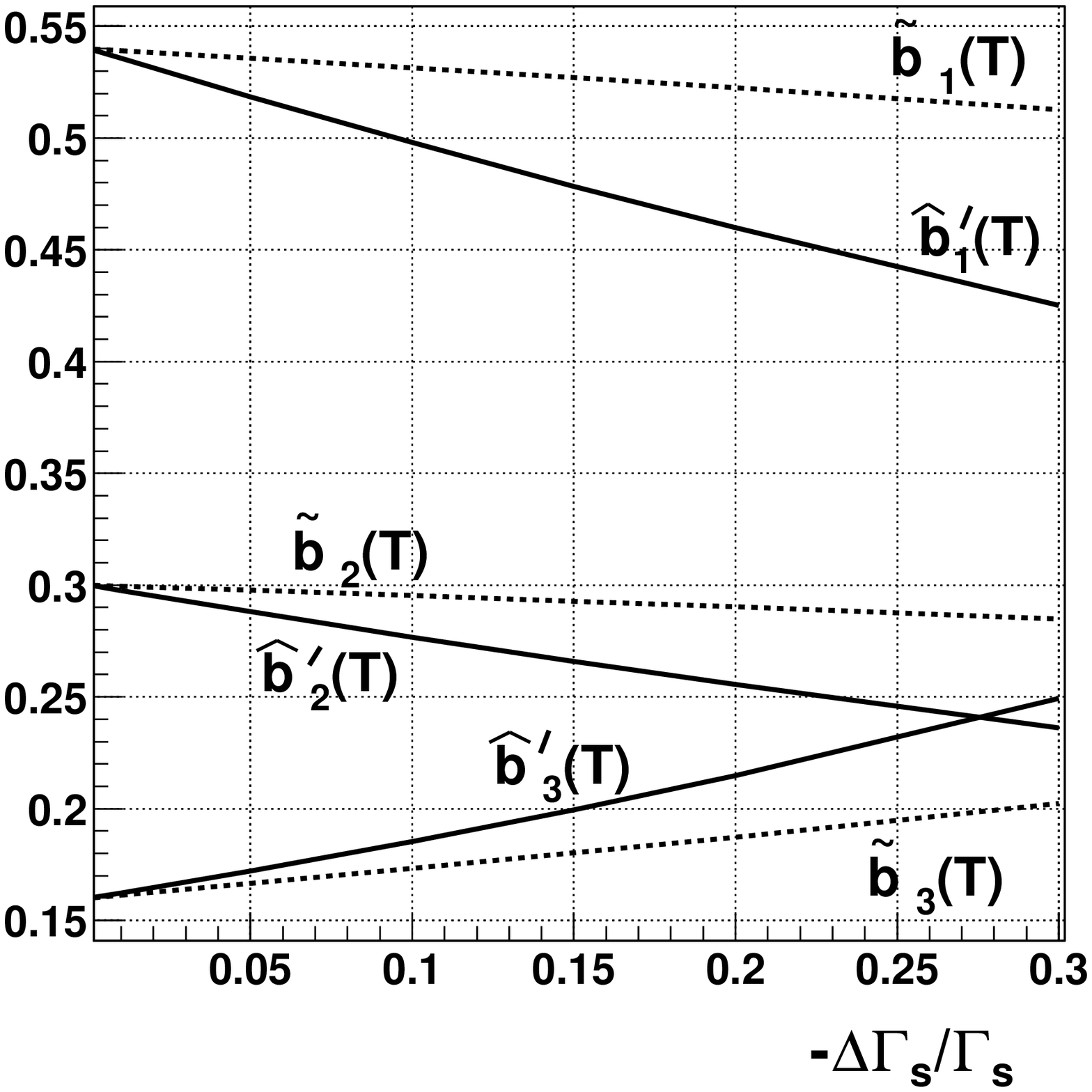}
\end{center}

{\bf Figure 4}

\vspace{5mm}
%=============================================================================

%=============================================================================
%                           Figure 5
%==========================================================================
\begin{center}
\epsfysize=0.35\textheight
\epsfbox{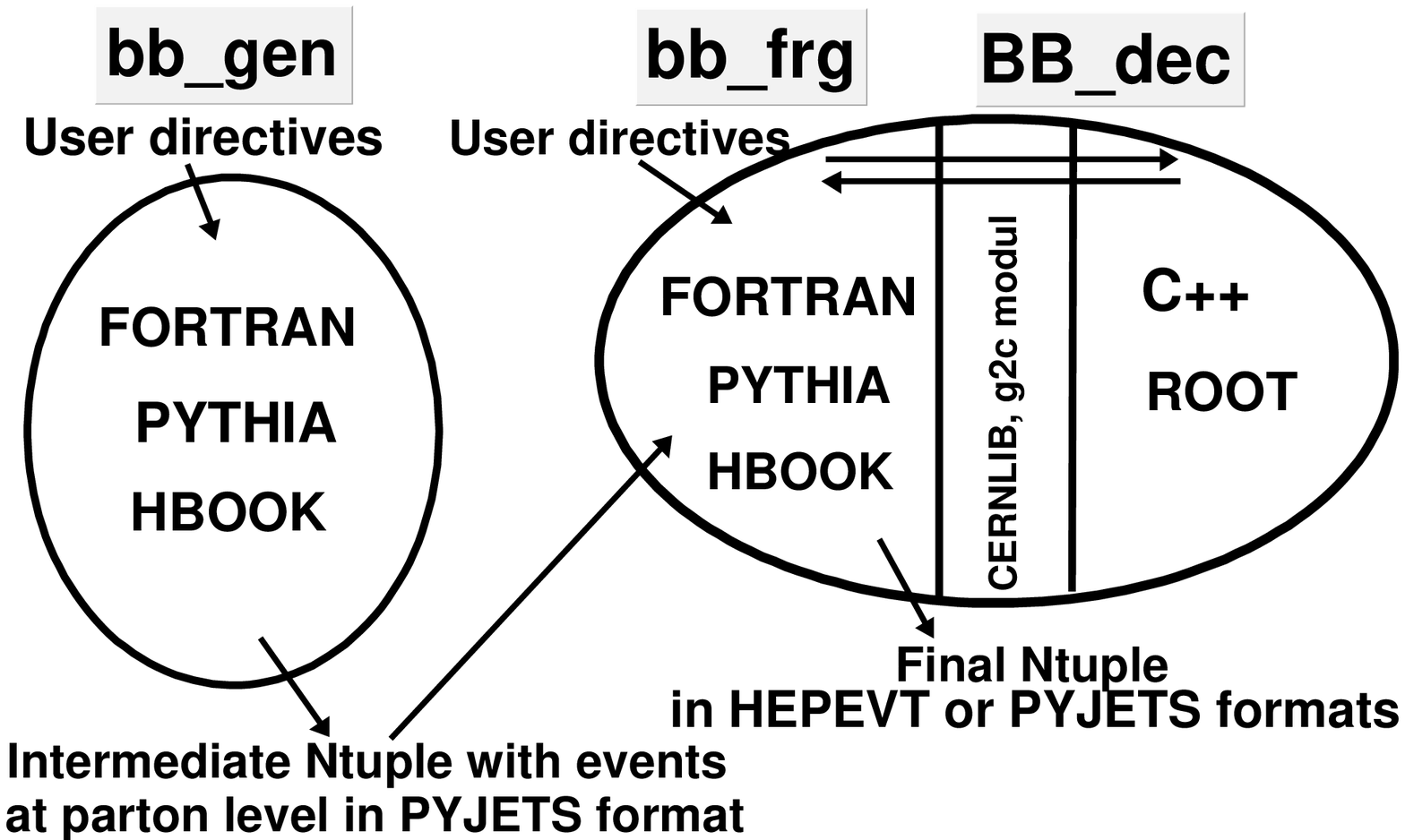}
\end{center}

\vspace{10mm}
{\bf Figure 5}
%==========================================================================

\end{document}